\documentclass
[twocolumn,superscriptaddress,secnumarabic,amssymb,amsmath,nobibnotes,aps,prd,showpacs,nofootinbib]{revtex4}%
\usepackage{graphicx}
\usepackage{epsf}
\usepackage{bm}
\usepackage{amsmath}
\usepackage{amsfonts}
\usepackage{amssymb}
\usepackage{epstopdf}
\usepackage{natbib}
\usepackage{rotating}
\usepackage{pdflscape}
\usepackage{adjustbox}
\usepackage{color}%
\providecommand{\U}[1]{\protect\rule{.1in}{.1in}}
\newcommand{\be}{\begin{equation}}
\newcommand{\ee}{\end{equation}}

\newcommand{\mincir}{\raise
-3.truept\hbox{\rlap{\hbox{$\sim$}}\raise4.truept\hbox{$<$}\ }}
\newcommand{\magcir}{\raise
-3.truept\hbox{\rlap{\hbox{$\sim$}}\raise4.truept\hbox{$>$}\ }}

\begin{document}
\title{Forecast constraints on Anisotropic Stress in Dark Energy using gravitational-waves}

\author{Weiqiang Yang}
\email{d11102004@163.com}
\affiliation{Department of Physics, Liaoning Normal University, Dalian, 116029, P. R.
China}

\author{Supriya Pan}
\email{supriya.maths@presiuniv.ac.in}
\affiliation{Department of Mathematics, Presidency University, 86/1 College Street, Kolkata 700073, India}

\author{David F. Mota}
\email{mota@astro.uio.no}
\affiliation{Institute of Theoretical Astrophysics, University of Oslo, 0315 Oslo, Norway}

\author{Minghui Du}
\email{angelbeats@mail.dlut.edu.cn}
\affiliation{Institute of Theoretical Physics, School of Physics, Dalian University of Technology, Dalian, 116024, P. R. China}

\pacs{98.80.-k, 95.36.+x, 95.35.+d, 98.80.Es}

\begin{abstract}
It is always interesting to investigate how well can a future experiment perform with respect to others (present or future ones). Cosmology is really an exciting field where a lot of puzzles are still unknown. In this article we consider a generalized dark energy (DE) scenario where anisotropic stress is present. We constrain this generalized cosmic scenario with an aim to investigate how gravitational waves standard sirens (GWSS) may constrain the anisotropic stress, which according to the standard cosmological probes, remains unconstrained.  In order to do this, we generate the luminosity distance measurements from $\mathcal{O} (10^3)$ mock GW events which match the expected sensitivity of the Einstein Telescope. 
Our analyses report that, first of all, GWSS can give better constraints on various cosmological parameters compared to the usual cosmological probes, but the viscous sound speed appearing due to the dark energy anisotropic stress, is totally unconstrained even after the inclusion of GWSS.   
\end{abstract}
\maketitle

%%%%%%%%%%%%%%

%%%%%%%%%%%%%%%%%%%%%%%%%%%%%%%%%%%%%%%%%%%%%%%%%%%%%%%%%%%%%%%%%%
%%%%%%%%%%%%%%%%%%%%%%%%%%%%%%%%%%%%%%%%%%%%%%%%%%%%%%%%%%%%%%%%%%
\section{Introduction}

The thrilling chapter of modern cosmology begun with the 
late-time accelerating phase of our universe \cite{Ade:2015xua}. 
This accelerated expansion is usually ascribed by 
the introduction of some exotic fluid with high negative pressure. This
exotic fluid could be either some dark energy fluid (in the context
of Einstein's general theory of gravity) \cite{Copeland:2006wr} or some geometrical dark energy (coming from the modified gravity theories) 
\cite{Nojiri:2010wj,Capozziello:2011et,Nojiri:2017ncd}. 
In this article we confine ourselves into the first approach,
that means, dark energy fluid.  In other words, our discussions will be 
restricted to the Einstein's gravitational theory. 
Within the framework of general relativity,   
a cluster  of dark energy models (see \cite{Copeland:2006wr} for various dark energy models) have been introduced in the literature, however, most of them are basically the variations of either the cosmological constant or some scalar field theory \cite{Wetterich:1987fm}. 
These variations naturally include different kind of couplings between the matter components of the universe \cite{Amendola:1999er,Mota:2006ed,Mota:2006fz}, couplings to the gravitational sector \cite{Koivisto:2006xf,Koivisto:2006ai}, or some non-canonical scalar field models, such as tachyons \cite{Sen:2002nu,Sen:2002in,Sen:2002an,Sen:2003mv,Sen:2002qa,Gibbons:2002md}, K-essence \cite{ArmendarizPicon:2000ah} etc. All such models are usually directed to explain the late accelerated expansion of the universe.  
Having a number of different dark energy models, it is quite natural to examine some important features of the dark energy models that might be able to discriminate between them, or alternatively, such features might be able to provide some potential techniques that may act as a baseline to construct the dark energy models. Certainly, the searching of these features are indeed worth exploring based on the current cosmological 
research.

However, it has been  already explored that in order to rule out some dark energy models or to discriminate between a number of existing dark energy models, the background evolution is not  sufficient at all. The evolution of the models at the level of perturbations must be considered into the picture in order to have a wider understanding of the models. 
According to the theory of general relativity, 
any matter component with equation of state $w \neq -1$, where $w=-1$ denotes the 
cosmological constant, must fluctuate. Thus, it is straightforward to realize that any dark energy model with $w \neq -1$ should have perturbations.  However, it might happen that such perturbations could be small enough depending on the nature of the equation of state $w$. For example, if the underlying dark energy fluid is smooth enough or if the Jeans length of this dark energy is large, its perturbations might be restricted to very large scales only. This kind of feature is exhibited in  minimally coupled quintessence models because  for such models the sound speed of perturbations, $c_s^2$, is  equal to the speed of light, $c_s^2=1$, which  consequently sets a large Jeans length \cite{Bean:2003fb,Xia:2007km}. 
While on the other hand, for some  other  dark energy candidates, this does not usually happen  \cite{Bagla:2002yn,Mota:2004pa}, and as a consequence, one could differentiate between the dark energy models.

Moreover, aside from the equation of state 
$w$ and the sound speed of perturbations
$c_s^2$, one more important characteristic of a cosmic fluid is its anisotropic stress $\sigma$  \cite{Hu:1998kj}. 
Although for a class of cosmological models including 
minimally coupled scalar field and perfect fluids, the anisotropic stress vanishes, however, it is a generic property of realistic fluids with finite shear viscous coefficients \cite{Schimd:2006pa,Brevik:2004sd}. Let us note that while 
$w$ and $c_s^2$ respectively 
determine the background and perturbative evolutions of the underlying cosmic fluid 
that is rotationally invariant, the anisotropic stress $\sigma$ actually gives a quantification on  how much the pressure of the cosmic fluid varies with the direction. 
In fact, the  perturbation for anisotropic stress is very important for understanding the evolution of  inhomogeneities in the early universe \cite{Hu:1998kj,Koivisto:2005mm,Mota:2007sz}. Thus, undoubtedly it is a natural question to investigate whether the current 
observational data may indicate for a nonzero anisotropic stress perturbations
in the dark energy dominated late-accelerating universe.

The effects of anisotropic stress, that may appear due to possible viscosity of dark energy, have not been paid much attention in the literature.  The reason 
for neglecting the anisotropic stress is that, for 
the conventional dark energy fluids, such as the 
cosmological constant or canonical scalar field models, $\sigma=0$. 
But, that should not be a logical case since
there is no such fundamental theory available yet 
that could correctly describe the actual dynamics of dark energy,  hence, 
the assumption of $\sigma = 0$ for any dark energy model, does not have any sense anymore. Therefore, from an unbiased scientific point of view, 
the presence of an anisotropic stress into the cosmic sector should be fairly considered and it is better to examine its non-null character, if any, with the help of recently available observational data.  Some earlier analyses have shown that coupled scalar field models have a non-negligible anisotropic stress \cite{Schimd:2006pa}. Additionally, dark energy vector field candidates proposed in \cite{ArmendarizPicon:2004pm,Kiselev:2004py} also allow nonzero anisotropic stress. 
Furthermore,  some model independent analyses \cite{Pinho:2018unz,Arjona:2020kco} has strongly argued that a non-null anisotropic stress in DE should be present according to the recent observational data.  Therefore, it is fairly  clear that the generalized dark energy models including the anisotropic stress, should be investigated in detail and several investigators, see for instance \cite{Saltas:2010tt,Sapone:2013wda, Amendola:2013qna,Cardona:2014iba,Chang:2014mta,Chang:2014bea,Majerotto:2015bra} tried to model the anistropic stress in various ways as well as confronted it with the observational data.  In particular, the role of an anisotropic stress was  investigated  in an interacting DE-DM scenario \cite{Yang:2018ubt}.

In this article, our approach is very appealing. We want to probe the anisotropic stress  using the gravitational waves data, detected recently \cite{Abbott:2016blz,Abbott:2016nmj,Abbott:2017vtc,Abbott:2017gyy,Abbott:2017oio}. In particular, we shall use the simulated gravitational waves standard sirens (GWSS) data to constrain the anisotropic stress with an aim to what future cosmological probe can tell us. The simulated GWSS data have already proved its super constraining power applied recently to various cosmological models, see for instance  \cite{DiValentino:2017clw,DiValentino:2018jbh,Du:2018tia,Yang:2019bpr,Yang:2019vni}. In fact, gravitational waves data are believed to offer more information about the nature of dark matter, dark energy and modified gravity theories \cite{Maselli:2016ekw,Bettoni:2016mij,Cai:2016hqj,Baker:2017hug,Creminelli:2017sry,Ezquiaga:2017ekz,Oost:2018tcv,Casalino:2018tcd, Zhao:2018gwk, Liu:2018sia,Chakraborty:2017qve,Visinelli:2017bny,Addazi:2017nmg,Flauger:2017ged, Cai:2018rzd}. In the recent past the effects of the GWSS data on various cosmological theories and related key parameters have been greatly studied and due to its growing interest in the cosmological community as reflected in  a series of investigations performed by many investigators \cite{Wei:2018cov,Zhang:2018dxi,Kase:2018aps,Lin:2018ken,Nunes:2018evm,Copeland:2018yuh,Casalino:2018wnc,Shafieloo:2018qnc,Zhao:2019kif,Nunes:2019bjq,Belgacem:2019pkk,DAgostino:2019hvh,Qiao:2019wsh,Zhao:2019xmm,Zhao:2019szi,Bonilla:2019mbm}, it is clearly realized that GWSS are in the limelight of modern cosmology. Thus, the consideration  of GWSS has significant effects on the dynamics of our universe. 
In this article, we use the simulated GWSS data from the Einstein Telescope \cite{Punturo:2010zz} (see also a number of works focused on this specific telescope \cite{Sathyaprakash:2009xt,Gair:2010dx,Hannam:2009vt,ChassandeMottin:2010zh,Mishra:2010tp,Zhao:2010sz,Huerta:2010un,Huerta:2010tp,Regimbau:2012ir,Taylor:2012db,Sathyaprakash:2012jk,Cai:2016sby,Zhang:2017sym,Zhang:2019ple,Bachega:2019fki,Zhang:2019loq}), however, technically, 
one can equally consider other observatories like 
Laser Interferometer Space Antenna (LISA) \cite{Audley:2017drz}, Deci-hertz Interferometer Gravitational wave Observatory (DECIGO) \cite{Kawamura:2011zz}, TianQin \cite{Luo:2015ght}. In fact, it will be interesting to use different simulated GWSS data from all the above sources with an aim to a detailed investigations in this direction.

The work has been organized in the following manner. 
In section \ref{sec2} we review the 
parameterization of a generalized cosmic fluid and introduce its connection with anisotropic stress. In section \ref{sec-data} we describe the observational data, namely both standard cosmological data and the simulated gravitational waves data. Then in section \ref{sec-results} we describe the extracted results obtained after using varieties of observational (real and forecasted) datasets.  Finally, in section \ref{sec-conclu} we close the present work with the main findings and comment on some future works that should be performed along these lines.

\section{Parameterizing Dark energy stress: FLRW background}
\label{sec2}

The energy momentum tensor of a general cosmic fluid  is defined as \cite{Koivisto:2005mm,Mota:2007sz}
\begin{eqnarray}\label{general-fluid}
T_{\mu\nu}= \rho u_\mu u_\nu + ph_{\mu\nu} + \Sigma_{\mu\nu},
\end{eqnarray}
where $p$, $\rho$ are respectively the pressure and energy density of 
the perfect fluid; 
$u_\mu$ is the four-velocity vector of this fluid, and $h_{\mu\nu}$, the projection tensor is defined as, $h_{\mu\nu} \equiv g_{\mu\nu} + u_\mu u_\nu$.
The quantity $\Sigma_{\mu\nu}$ in (\ref{general-fluid}) may include only spatial inhomogeneity which vanishes for a perfect fluid, that means, 
$\Sigma_{\mu\nu} \equiv 0$. Additionally, 
for a homogeneous and isotropic universe, $\Sigma_{\mu\nu}$ is also zero at the level of background; in fact, in  such a case, it denotes the anisotropic perturbation at the first order.
Thus, at the background level, since $\Sigma_{\mu\nu} =0$, the evolution of the fluid (\ref{general-fluid}) 
is determined by the continuity equation,
\begin{eqnarray}\label{continuity}
\dot{\rho} + 3H (p+ \rho) = 0.
\end{eqnarray}
where $H$ is the Hubble rate of the Friedmann-Lema\^{i}tre-Robertson-Walker (FLRW) universe. The condition for the adiabaticity  
of a fluid is $ p = p (\rho)$, which tells us that the evolution of the
sound speed is determined by the equation of state $w = p/\rho$, alone. However, being the most general,   
the sound speed is defined as the ratio
of pressure to the density perturbations in the frame comoving
with the dark energy fluid, defined as 

\begin{eqnarray}
c^2_{s,a}\equiv\frac{\dot{p}}{\dot{\rho}} = w-\frac{\dot{w}}{3\mathcal{H}(1+w)},
\end{eqnarray}
where an overhead dot represents the differentiation with respect to the conformal time $\tau$;  
$w$ being the equation of state defined as 
$w\equiv p/\rho$; $\mathcal{H}=\dot{a}/a$ is the conformal Hubble parameter. We note that the relation between the perturbations of $\delta p$ and $\delta\rho$ is, $\delta p=c^2_{s,a}\delta \rho$. However, for an entropic fluid, the pressure $p$ may not only depend on the energy density $\rho$. In fact, there might have another degree of freedom in order to describe the microphysical properties of the general cosmic fluid and such microphysical property is usually encoded in the effective speed of sound $c^2_{s,\rm eff}$ defined as 
\begin{equation}
c^2_{s,\rm eff}\equiv \frac{\delta p}{\delta \rho}|_{\rm rf},
\end{equation}   
in the rest-frame (`{\rm rf}') of the underlying cosmic fluid. In absence of entropic perturbation, $c^2_{s,\rm eff}=c^2_{s,a}$. Therefore, one may easily conclude that a perfect fluid is completely characterized  by two quantities, one is its equation of state $w$ and the other is its effective speed of sound $c^2_{s,\rm eff}$.

However, aside from the previous quantities, namely, $w$ and 
$c^2_{s,\rm eff}$ associated with a cosmic fluid, one more important quantity is needed in order to understand the cosmic fluid at the level of background and perturbations. This quantity is the anisotropic stress $\sigma$, and it should  be considered into the cosmological framework even in an isotropic and homogeneous FLRW universe, where the anisotropic stress $\sigma$ maybe taken as the  spatial perturbation. 
This anisotropic stress actually distinguishes  between the Newtonian potential and curvature perturbation in the conformal Newtonian gauge.

So, now one can calculate the evolution  equations at the level of perturbations for the above model considering any gauge. We choose the synchronous gauge in this article and using this gauge, the density perturbations and velocity perturbations can be written as  \cite{Ma:1995ey}
\begin{eqnarray}
\dot{\delta}&=&-(1+w)\left(\theta+\frac{\dot{h}}{2}\right)-3 \mathcal{H}\left(\frac{\delta p}{\delta \rho}-w \right)\delta,\label{eq:continue}\\
\dot{\theta}&=&-\mathcal{H}\left(1-3c^2_{s,a} \right)+\frac{\delta p/\delta \rho}{1+w}k^{2}\delta-k^{2}\sigma,\label{eq:euler}
\end{eqnarray}
where the anisotropic stress $\sigma$ is related  to $\Sigma_{\mu\nu}$ (see eqn. \ref{general-fluid}) via $(\rho+p)\sigma\equiv -(\hat{k}_i\hat{k}_j-\delta_{ij}/3)\Sigma^{ij}$. Now, using the effective speed of sound, one can recast the above equations as 

% \begin{widetext}
 \begin{eqnarray}
 \dot{\delta}=-(1+w)\left(\theta+\frac{\dot{h}}{2}\right)+\frac{\dot{w}}{1+w}\delta \nonumber\\ -3\mathcal{H}(c^2_{s,\rm eff}-c^2_{s,a})\left[\delta + 3\mathcal{H}(1+w)\frac{\theta}{k^2}\right],\\
\dot{\theta}=-\mathcal{H} \left(1-3c^2_{s,\rm eff} \right)\theta + \frac{c^2_{s,\rm eff}}{1+w}k^2\delta-k^2\sigma,
 \end{eqnarray}
%\end{widetext}
where following Hu \cite{Hu:1998kj} we suppose that the anisotropic stress $\sigma$ satisfies the evolution equation 
\begin{eqnarray}
\dot{\sigma}+3\mathcal{H}\frac{c^2_{s,a}}{w}\sigma=\frac{8}{3}\frac{c^2_{vis}}{1+w}\left(\theta+\frac{\dot{h}}{2}+3\dot{\eta}\right),\label{eq:shear}
\end{eqnarray}
where $c^2_{vis}$ is the viscous speed of sound and it controls  the relation between velocity/metric shear and the anisotropic stress. For a relativistic fluid, $c^2_{vis} = 1/3$. For any dark energy fluid, $c^2_{vis}$ acts as a free model parameter to be determined by the observational data. In connection with that we recall a recent study \cite{Arjona:2018jhh} where the authors argue that $c^2_{vis}$ may have a dynamical nature, and this is an interesting issue which  should be further investigated in light of the presently available potential cosmological probes. In the present we shall however treat $c_{vis}^2$ to be a free parameter and following \cite{Huey:2004jz,Koivisto:2005mm}, one notices that the value of $c^2_{vis}/(1+w)$ should be positive. This forces us to consider to different models, namely, (i) the dark energy equation of state is in the quintessence regime, that means, 
$-1<w < 0$ (Model I) and (ii) when the dark energy equation of state crosses the phantom divide line, that means, $w < -1$ (Model II). Let us note that for all the analyses that will be described in section \ref{sec-results}, we adopted the adiabatic initial conditions similar to the authors in Ref. \cite{Koivisto:2005mm}.

\section{Standard Cosmological Probes and the GWSS data}
\label{sec-data}

We now proceed to extract the cosmological constraints using a set of usual dark energy probes and the simulated gravitational waves standard sirens (GWSS). In this section we shall describe the standard cosmological data and then we will refer some works that describe the methodology to generate the mock GWSS. In what follows we describe the standard cosmological probes first.

\begin{enumerate}

\item {\bf CMB from Planck 2015:} The cosmic microwave background (CMB) data are one of the potential data to probe the nature of dark energy. We use the Planck 2015 measurements \cite{Adam:2015rua, Aghanim:2015xee} that include the high- and low- $\ell$ TT likelihoods (in the multipole range $ 2 \leq \ell \leq 2508$) plus the high- and low- $\ell$ polarization likelihoods. In the article we refer this dataset as P15. 

\item {\bf CMB from Planck 2018:} We also consider the latest CMB data from Planck 2018 final release \cite{Aghanim:2018oex,Aghanim:2019ame}. In the article we refer this dataset as P18. 

\item {\bf BAO:} We include the baryon acoustic oscillations (BAO) distance measurements from different measurements \cite{Beutler:2011hx, Ross:2014qpa,Gil-Marin:2015nqa}.

\item {\bf SNIa:} We use the latest compilation of Pantheon sample from Supernovae Type Ia (SNIa) \cite{Scolnic:2017caz}). We note that Supernovae Type Ia data were the first astronomical data that signaled for the existence of some hypothetical dark energy fluid in the universe's sector. Thus, this particular data play a very important role in dark energy analysis.

\end{enumerate}

The methodology to generate the GWSS is described by the present authors in  \cite{Du:2018tia} (also see \cite{Yang:2019bpr}). The present work deals with the same methodology as in \cite{Du:2018tia,Yang:2019bpr}, thus, we avoid the repetition here and directly refers to \cite{Du:2018tia,Yang:2019bpr} for details. 
We would mark some important points here. Using the methodology here for this work we generate 1000 simulated GW data for our purpose. The generation of 1000 GW data is as follows. We first  constrain the cosmological scenarios using the usual cosmological probes, such as CMB, BAO, Pantheon. Then we use the best-fit values of all the free and derived parameters obtained from the standard cosmological probes and assuming the present anisotropic dark energy models as the fiducial models, and following exactly similar technique described in \cite{Du:2018tia,Yang:2019bpr} we generate the 1000 mock GW data. Next we add these 1000 mock GW data to these standard cosmological probes and constrain the underlying cosmological scenarios. For the entire statistical analysis we use the markov chain monte carlo package \textit{cosmomc} \cite{Lewis:2002ah, Lewis:2013hha} equipped with a convergence diagnostic by Gelman-Rubin \cite{Gelman-Rubin}. The \texttt{cosmomc} code also supports the Planck 2015 Likelihood \cite{Aghanim:2015xee} and  Planck 2018 likelihood \cite{Aghanim:2018oex,Aghanim:2019ame}  \footnote{See \url{http://cosmologist.info/cosmomc/.}, a freely available code to extract the cosmological constraints.}. 
Finally, in Table \ref{tab:priors} we enlist the flat priors on the cosmological parameters used during the time of statistical analysis.  
\begin{table}
\begin{center}
\begin{tabular}{|c|c|c|c}
\hline 
Parameter   & Prior (Model I) & Prior (Model II) \\
\hline 
$\Omega_{b} h^2$             & $[0.005,0.1]$ & $[0.005,0.1]$\\
$\Omega_{c} h^2$             & $[0.01,0.99]$ & $[0.01,0.99]$\\
$\tau$                       & $[0.01,0.8]$ & $[0.01,0.8]$ \\
$n_s$                        & $[0.5, 1.5]$ & $[0.5, 1.5]$\\
$\log[10^{10}A_{s}]$         & $[2.4,4]$ & $[2.4,4]$ \\
$100\theta_{MC}$             & $[0.5,10]$ & $[0.5,10]$\\ 
$w$                          & $[-1, 0]$ & $[-3, -1]$\\
$c_{vis}^2$                  & $[0, 10]$  & $[-10, 0]$\\
\hline 
\end{tabular}
\end{center}
\caption{Flat priors set on various cosmological parameters for the statistical analysis. }
\label{tab:priors}
\end{table}
\begingroup                                                                                                                     
%\squeezetable                                                                                                                   
\begin{center}                                                                                                                  
\begin{table*}
\scalebox{0.9}
{                                                                                                                                                                                                                                      
\begin{tabular}{cccccccccccccccc}                                                                                                            
\hline\hline                                                                                                                    
Parameters & P15 & P15+BAO & P15+Pantheon & P15+BAO+Pantheon\\ \hline
$\Omega_c h^2$ & $    0.1194_{-    0.0015-    0.0028}^{+    0.0014+    0.0029}$ & $    0.1177_{-    0.0011-    0.0022}^{+    0.0012+    0.0021}$ & $    0.1186_{-    0.0012-    0.0026}^{+    0.0014+    0.0025}$ & $    0.1178_{-    0.0010-    0.0019}^{+    0.0010+    0.0019}$ \\

$\Omega_b h^2$ & $    0.02223_{-    0.00016-    0.00030}^{+    0.00015+    0.00031}$ & $    0.02236_{-    0.00014-    0.00028}^{+    0.00014+    0.00029}$ & $    0.02230_{-    0.00015-    0.00029}^{+    0.00015+    0.00030}$ & $    0.02236_{-    0.00014-    0.00028}^{+    0.00014+    0.00029}$ \\

$100\theta_{MC}$ & $    1.04073_{-    0.00033-    0.00066}^{+    0.00033+    0.00065}$ & $    1.04097_{-    0.00030-    0.00059}^{+    0.00030+    0.00059}$  & $    1.04083_{-    0.00031-    0.00061}^{+    0.00031+    0.00063}$ & $    1.04096_{-    0.00030-    0.00060}^{+    0.00030+    0.00058}$ \\

$\tau$ & $    0.081_{-    0.017-    0.034}^{+    0.017+    0.033}$ & $    0.089_{-    0.017-    0.032}^{+    0.016+    0.032}$ & $    0.083_{-    0.017-    0.033}^{+    0.017+    0.034}$ & $    0.087_{-    0.016-    0.032}^{+    0.016+    0.033}$  \\

$n_s$ & $    0.9654_{-    0.0046-    0.0090}^{+    0.0045+    0.0087}$ & $    0.9702_{-    0.0039-    0.0075}^{+    0.0040+    0.0077}$ & $    0.9677_{-    0.0048-    0.0088}^{+    0.0042+    0.0090}$ & $    0.9700_{-    0.0038-    0.0076}^{+    0.0037+    0.0075}$ \\

${\rm{ln}}(10^{10} A_s)$ & $    3.095_{-    0.033-    0.065}^{+    0.033+    0.063}$ & $    3.108_{-    0.033-    0.063}^{+    0.032+    0.063}$ & $    3.098_{-    0.033-    0.066}^{+    0.033+    0.066}$ & $    3.104_{-    0.032-    0.064}^{+    0.032+    0.064}$ \\

$w$ & $  < -0.854 < -0.737$ & $   < -0.953 < -0.904$ & $  < -0.973 < -0.942$ & $  < -0.974 < -0.946 $ \\

$c^2_{vis}$ & $\mbox{unconstrained}$ & $   \mbox{unconstrained}$ & $\mbox{unconstrained}$ & $\mbox{unconstrained}$ \\

$\Omega_{m0}$ & $    0.348_{-    0.037-    0.048}^{+    0.018+    0.060}$ & $    0.314_{-    0.009-    0.016}^{+    0.007+    0.017}$ & $    0.315_{-    0.009-    0.017}^{+    0.009+    0.018}$ & $    0.309_{-    0.006-    0.013}^{+    0.006+    0.013}$ \\

$\sigma_8$ & $    0.797_{-    0.022-    0.054}^{+    0.032+    0.049}$ & $    0.820_{-    0.015-    0.031}^{+    0.017+    0.029}$ & $    0.823_{-    0.014-    0.028}^{+    0.014+    0.028}$ & $    0.824_{-    0.014-    0.028}^{+    0.014+    0.027}$ \\

$H_0$ & $   64.11_{-    1.70-    4.93}^{+    3.22+    4.16}$ & $   66.98_{-    0.66-    1.72}^{+    0.97+    1.54}$ & $   67.09_{-    0.67-    1.55}^{+    0.80+    1.40}$ & $   67.51_{-    0.52-    1.16}^{+    0.61+    1.10}$ \\

\hline\hline                                                                                                                    
\end{tabular}
}                                                                                                                   
\caption{The table displays the constraints on various free and derived cosmological parameters at 68\% and 95\% CL for Model I using the usual cosmological probes, namely, P15, BAO and 
Pantheon. For the dark energy equation of state we present its upper limits at 68\% and 95\% CL. }
\label{tab:results1}                                                                                                   
\end{table*}                                                                                                                     
\end{center}                                                                                                                    
\endgroup  
\begin{figure*}
\includegraphics[width=0.38\textwidth]{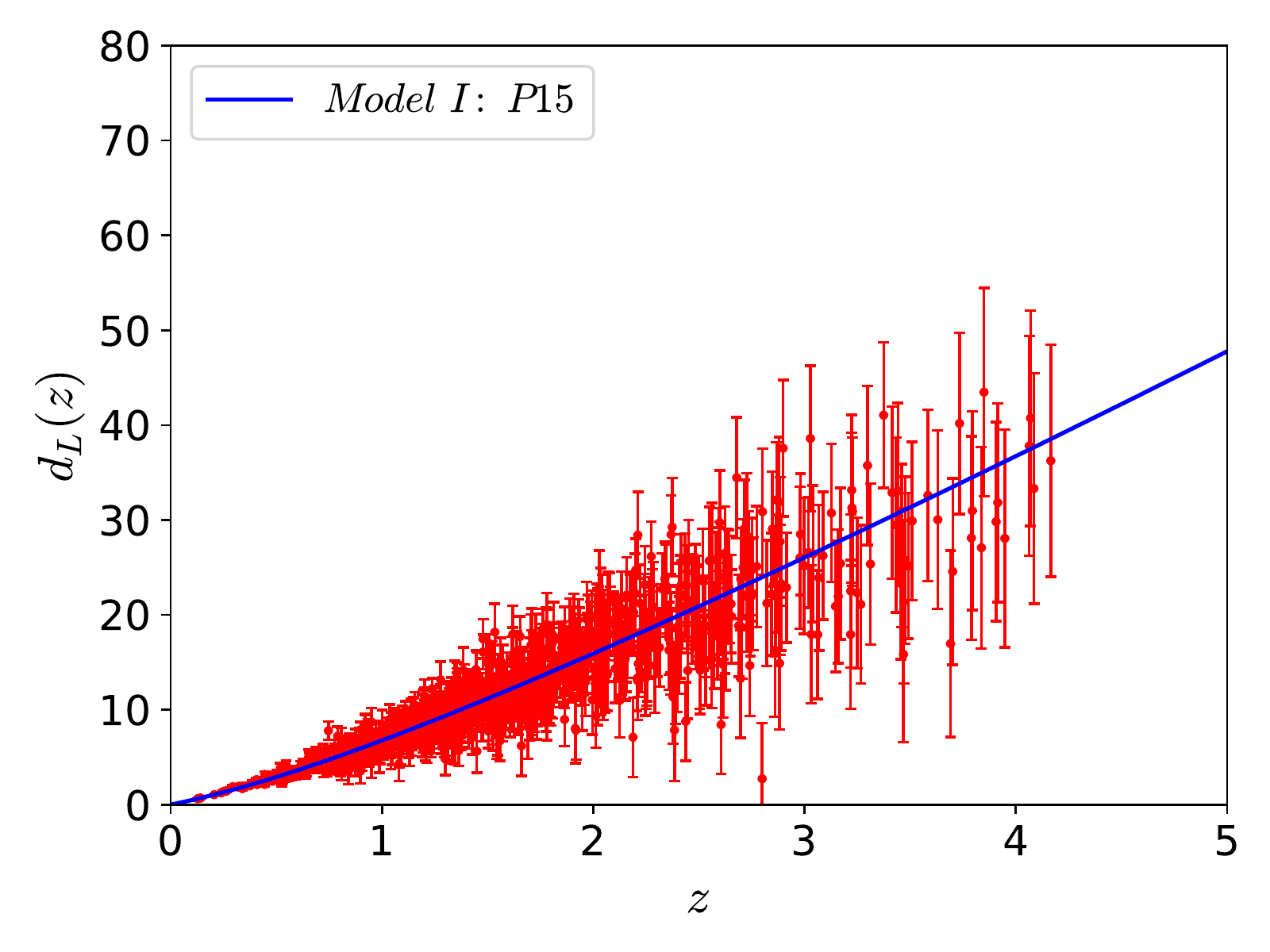}
\includegraphics[width=0.38\textwidth]{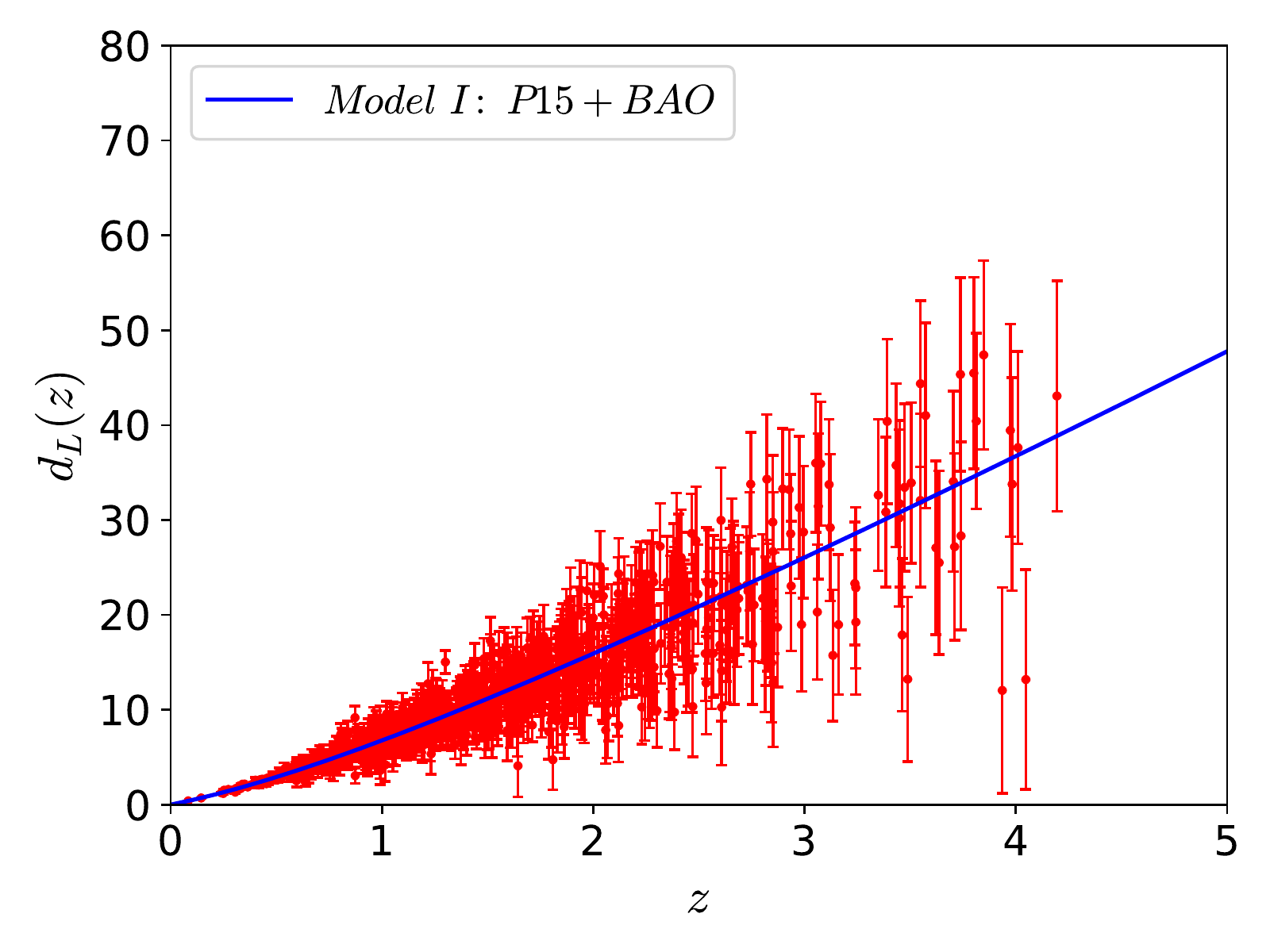}
\includegraphics[width=0.38\textwidth]{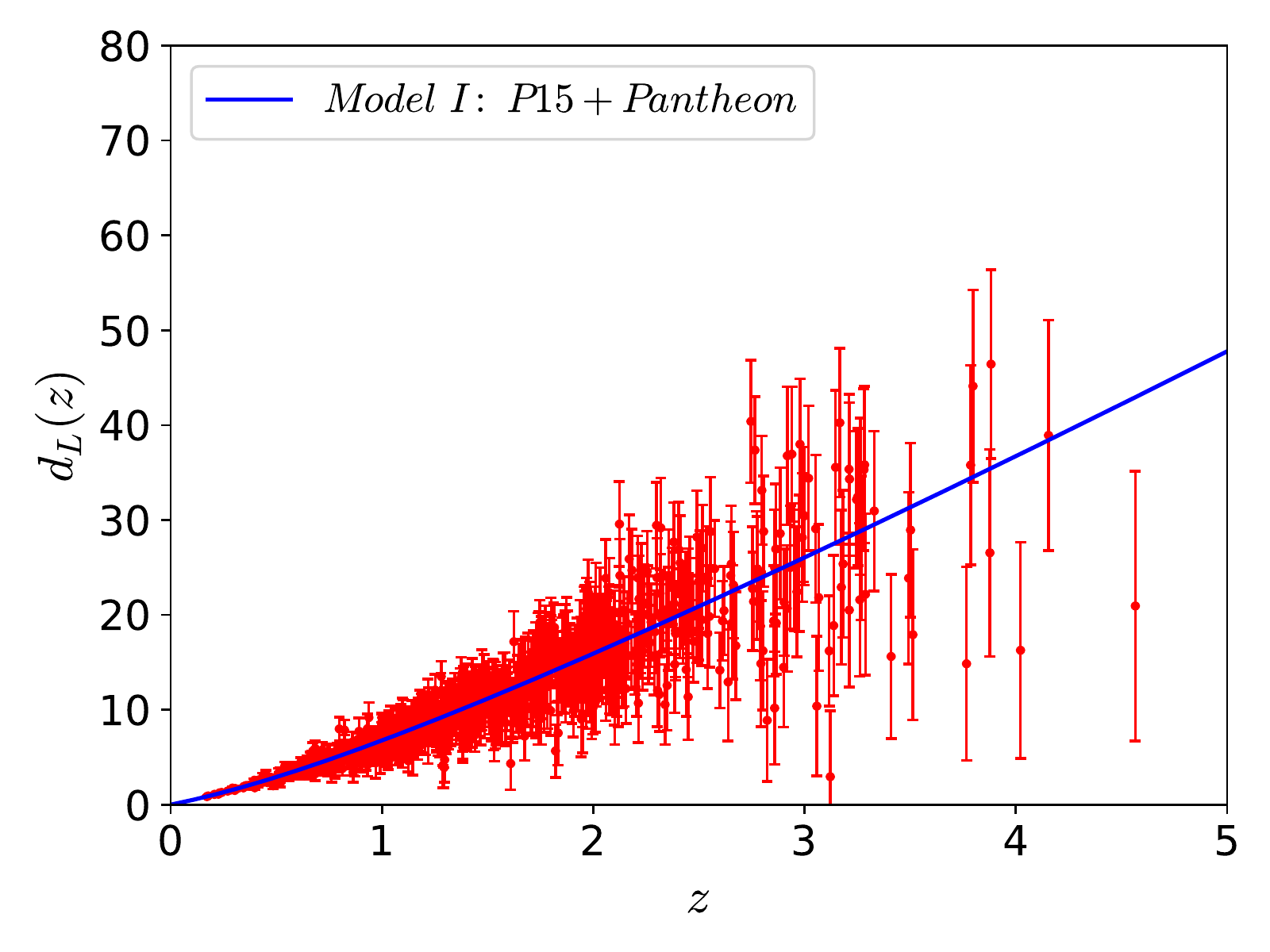}
\includegraphics[width=0.38\textwidth]{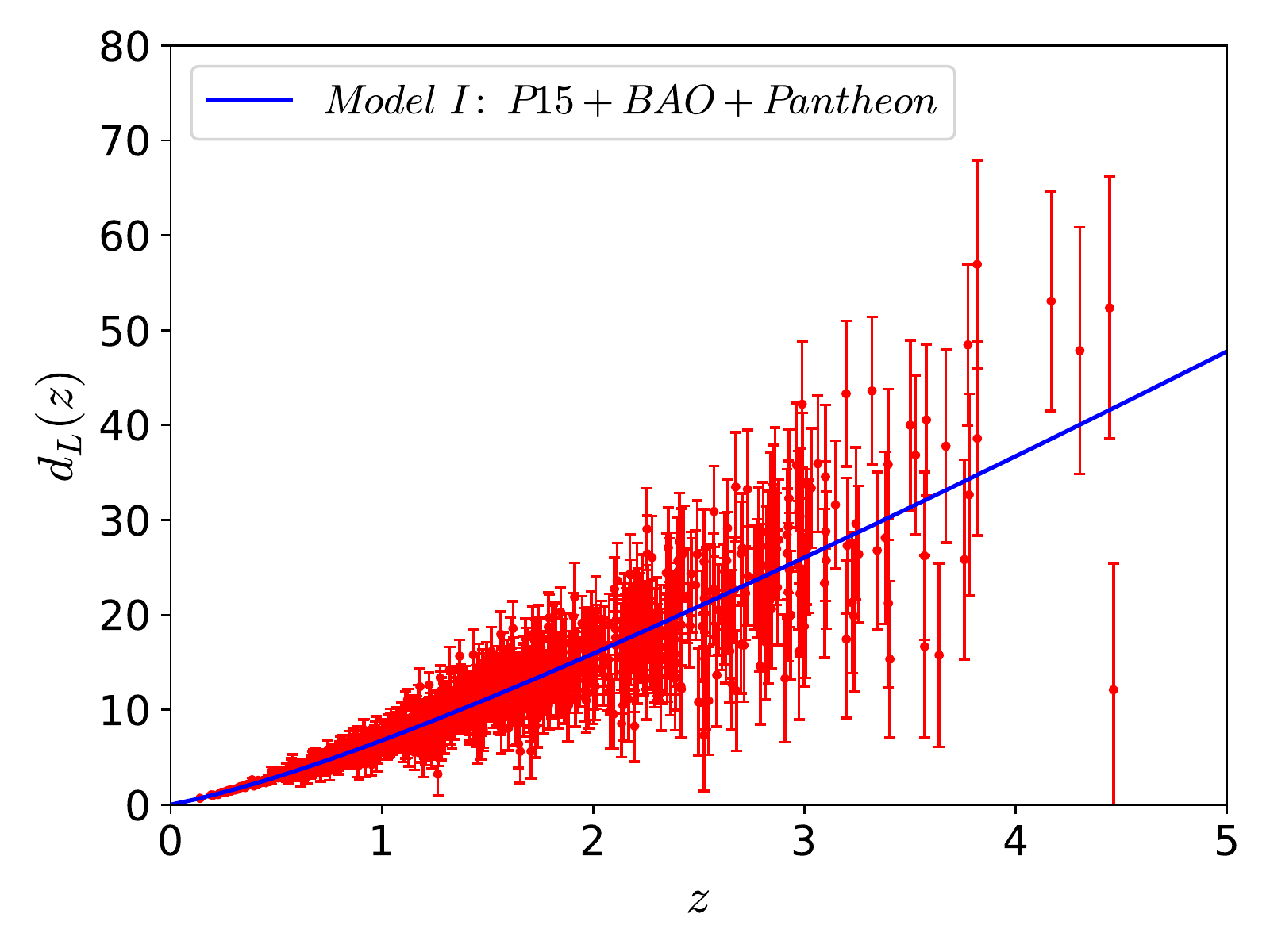}
\caption{For the fiducial (Model I) model we first constrain the cosmological parameters using the datasets P15, P15+BAO, P15+Pantheon and P15+BAO+Pantheon and then we use the best-fit of the parameters for ``each dataset'' to generate the corresponding GW catalogue. Following this, in each panel we show $d_L (z)$ vs $z$ catalogue with the corresponding error bars for 1000 simulated GW events. The upper left and upper right panels respectively present the catalogue ($z$, $d_L (z)$) with the corresponding error bars for 1000 simulated events derived using the P15 alone and P15+BAO dataset. The lower left and lower right panels respectively present the catalogue ($z$, $d_L (z)$) with the corresponding error bars for 1000 simulated events derived using the P15+Pantheon and P15+BAO+Pantheon datasets. }
\label{dl-modelI}
\end{figure*}
\begingroup                                                                                                                     
%\squeezetable                                                                                                                   
\begin{center}                                                                                                                  
\begin{table*}
\scalebox{0.9}
{                                                                                                                   
\begin{tabular}{ccccccccccccc}                                                                                                            
\hline\hline                                                                                                                    
Parameters & P15+GW & P15+BAO+GW & P15+Pantheon+GW &  P15+BAO+Pantheon+GW\\ \hline

$\Omega_c h^2$ & $    0.1190_{-    0.0010-    0.0020}^{+    0.0010+    0.0018}$ & $    0.1180_{-    0.0011-    0.0023}^{+    0.0012+    0.0021}$ & $    0.1181_{-    0.0009-    0.0019}^{+    0.0010+    0.0017}$ & $    0.1178_{-    0.0009-    0.0019}^{+    0.0010+    0.0018}$ \\

$\Omega_b h^2$ & $    0.02226_{-    0.00013-    0.00024}^{+    0.00012+    0.00025}$ & $    0.02233_{-    0.00014-    0.00027}^{+    0.00014+    0.00028}$ & $    0.02234_{-    0.00012-    0.00024}^{+    0.00012+    0.00025}$  & $    0.02235_{-    0.00013-    0.00025}^{+    0.00012+    0.00026}$ \\

$100\theta_{MC}$ & $    1.04078_{-    0.00030-    0.00058}^{+    0.00029+    0.00057}$ & $    1.04092_{-    0.00030-    0.00059}^{+    0.00031+    0.00059}$ & $    1.04092_{-    0.00030-    0.00057}^{+    0.00030+    0.00058}$  & $    1.04095_{-    0.00029-    0.00056}^{+    0.00029+    0.00056}$ \\

$\tau$ & $    0.082_{-    0.016-    0.031}^{+    0.016+    0.032}$ & $    0.087_{-    0.017-    0.035}^{+    0.017+    0.032}$ & $    0.086_{-    0.016-    0.032}^{+    0.018+    0.031}$ & $    0.088_{-    0.016-    0.033}^{+    0.018+    0.031}$ \\

$n_s$ & $    0.9667_{-    0.0036-    0.0069}^{+    0.0037+    0.0073}$ & $    0.9692_{-    0.0040-    0.0078}^{+    0.0040+    0.0082}$ & $    0.9691_{-    0.0036-    0.0068}^{+    0.0036+    0.0072}$ & $    0.9700_{-    0.0036-    0.0070}^{+    0.0036+    0.0073}$  \\

${\rm{ln}}(10^{10} A_s)$ & $    3.096_{-    0.032-    0.061}^{+    0.032+    0.063}$ & $    3.104_{-    0.033-    0.067}^{+    0.033+    0.064}$ & $    3.103_{-    0.032-    0.065}^{+    0.033+    0.062}$ & $    3.105_{-    0.031-    0.067}^{+    0.035+    0.062}$ \\

$w$ & $ < -0.974 < -0.948 $ & $  < -0.924 < -0.901 $  & $ < -0.979 < -0.957 $ & $ < -0.975 < -0.950 $ \\

$c^2_{vis}$ & $ \mbox{unconstrained}$ & $\mbox{unconstrained}$ & $    \mbox{unconstrained}$  & $\mbox{unconstrained}$ \\

$\Omega_{m0}$ & $    0.317_{-    0.005-    0.009}^{+    0.005+    0.009}$ & $    0.319_{-    0.006-    0.010}^{+    0.005+    0.011}$ & $    0.310_{-    0.004-    0.009}^{+    0.004+    0.009}$ & $    0.309_{-    0.004-    0.008}^{+    0.004+    0.008}$ \\

$\sigma_8$ & $    0.824_{-    0.014-    0.027}^{+    0.014+    0.027}$ & $    0.815_{-    0.015-    0.030}^{+    0.015+    0.030}$ & $    0.825_{-    0.014-    0.027}^{+    0.014+    0.026}$  & $    0.824_{-    0.014-    0.028}^{+    0.014+    0.027}$  \\

$H_0$ & $   66.94_{-    0.38-    0.80}^{+    0.39+    0.75}$ & $   66.45_{-    0.55-    1.16}^{+    0.63+    1.13}$ & $   67.48_{-    0.35-    0.73}^{+    0.39+    0.69}$ & $   67.52_{-    0.35-    0.78}^{+    0.42+    0.70}$ \\
\hline\hline                                                                                                                    
\end{tabular}
}                                                                                                                   
\caption{In this table we show the constraints on various free and derived cosmological parameters of Model I at 68\% and 95\% CL after the inclusion of GWSS data with the standard cosmological probes P15, BAO and Pantheon.  }
\label{tab:results1a}                                                                                                   
\end{table*}                                                                                                                     
\end{center}                                                                                                                    
\endgroup   
\begin{figure*}
    \centering
    \includegraphics[width=0.99\textwidth]{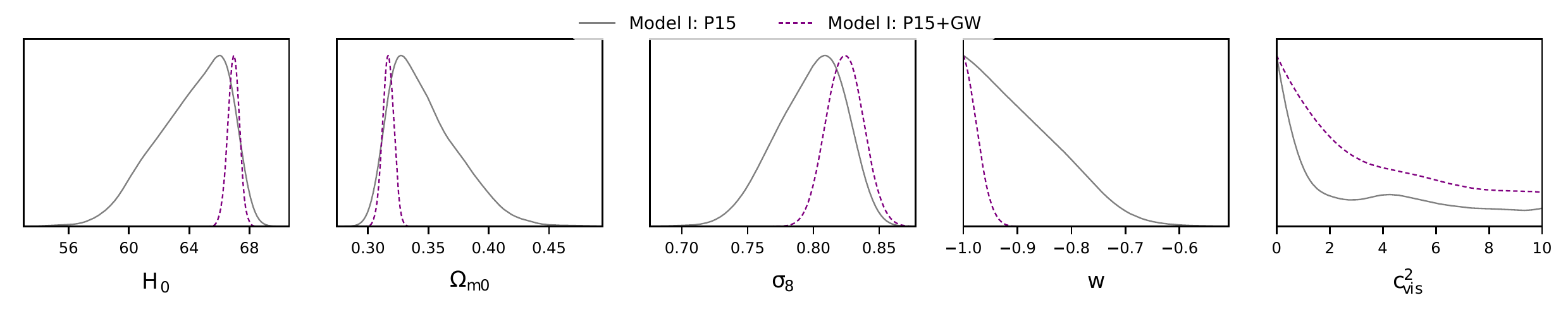}
    \caption{1-dimensional marginalized posterior distributions of some key parameters of Model I for the datasets P15 and P15+GW. }
    \label{fig:1D-ModelI}
\end{figure*}
\begin{figure*}
    \centering
    \includegraphics[width=0.39\textwidth]{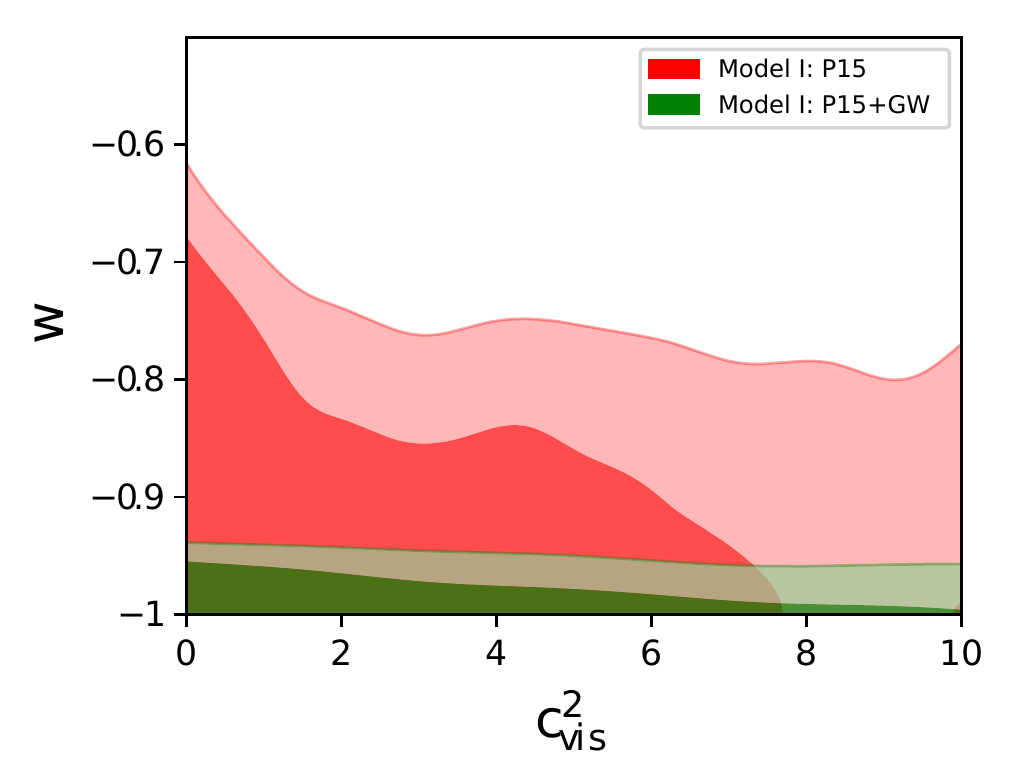}
    \includegraphics[width=0.38\textwidth]{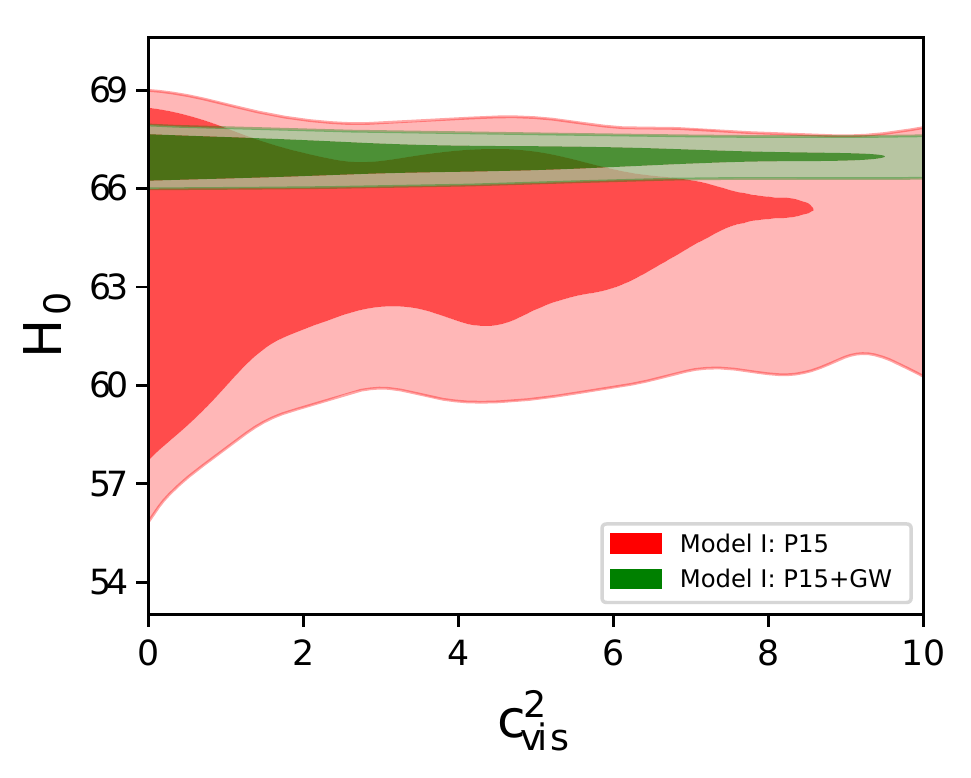}
    \includegraphics[width=0.38\textwidth]{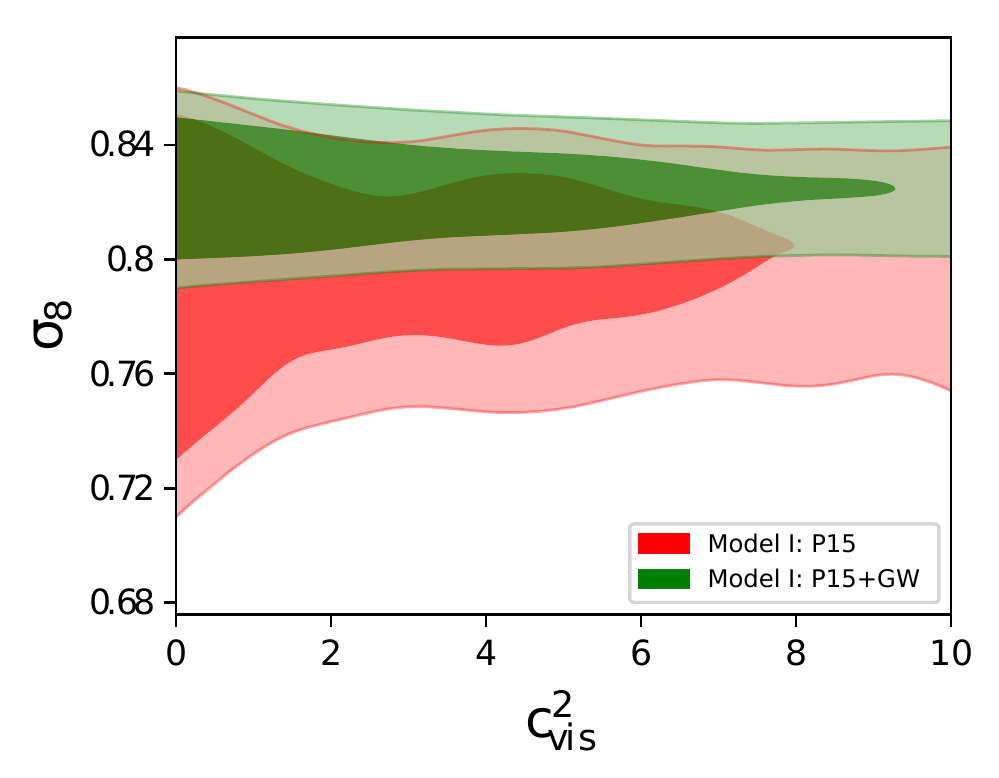}
     \includegraphics[width=0.38\textwidth]{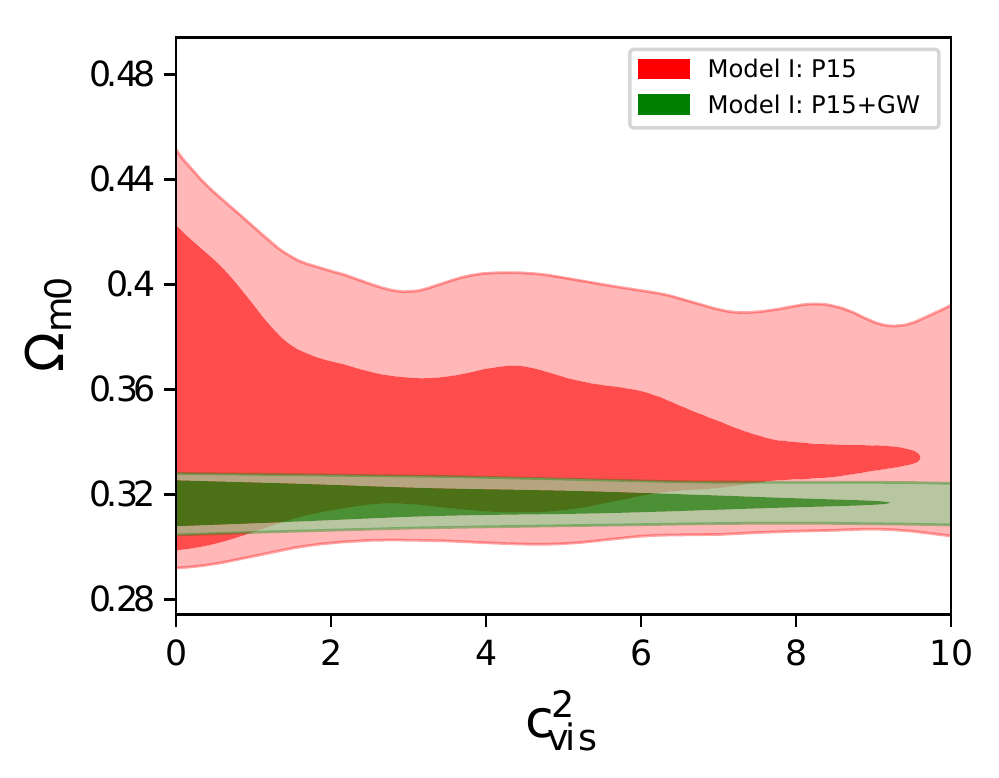}
    \caption{2-dimensional contour plots showing the dependence of $c_{vis}^2$ with other cosmological parameters of Model I for the P15 (red contours) and P15+GW (green contours) datasets. }
    \label{fig:2D-ModelI}
\end{figure*}

\section{Results}
\label{sec-results}

In this section we describe the main results on the cosmological parameters for the two variants of the model, namely, Model I ($w> -1$) and Model II ($w< -1$) using various observational datasets. In particular, we focus on the  effects of GW data on the cosmological parameters. 

\begin{figure*}
\includegraphics[width=0.38\textwidth]{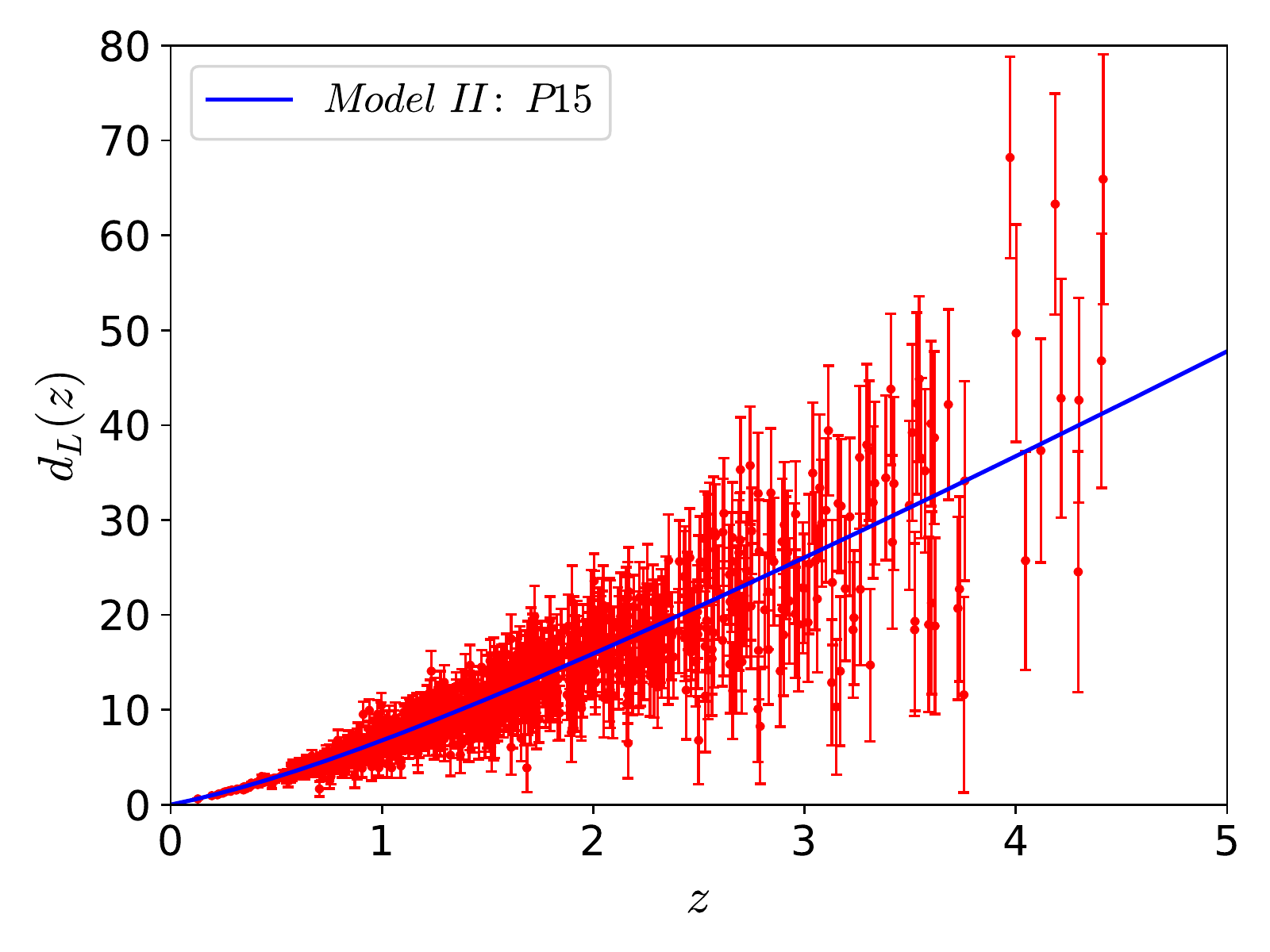}
\includegraphics[width=0.38\textwidth]{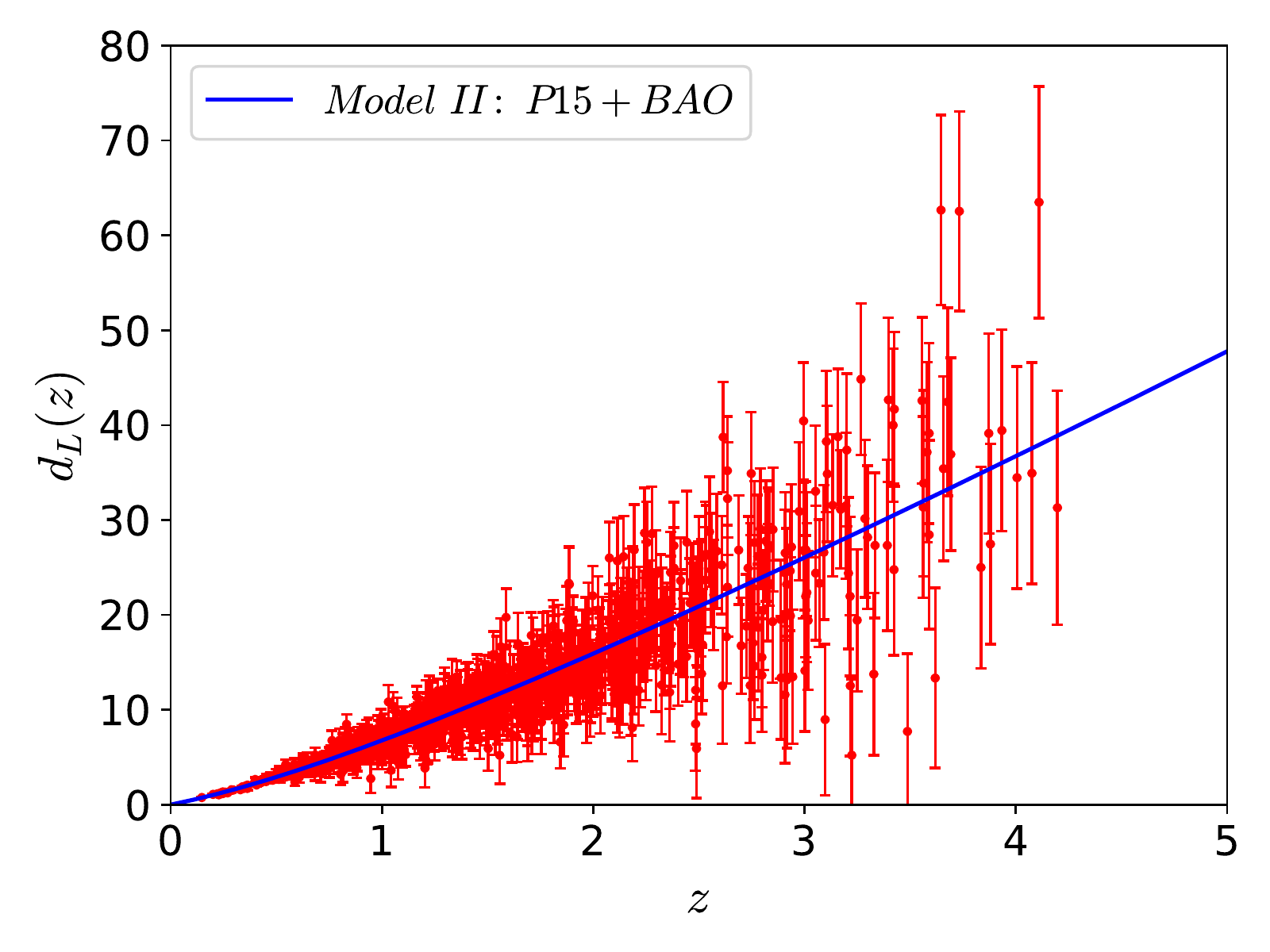}
\includegraphics[width=0.38\textwidth]{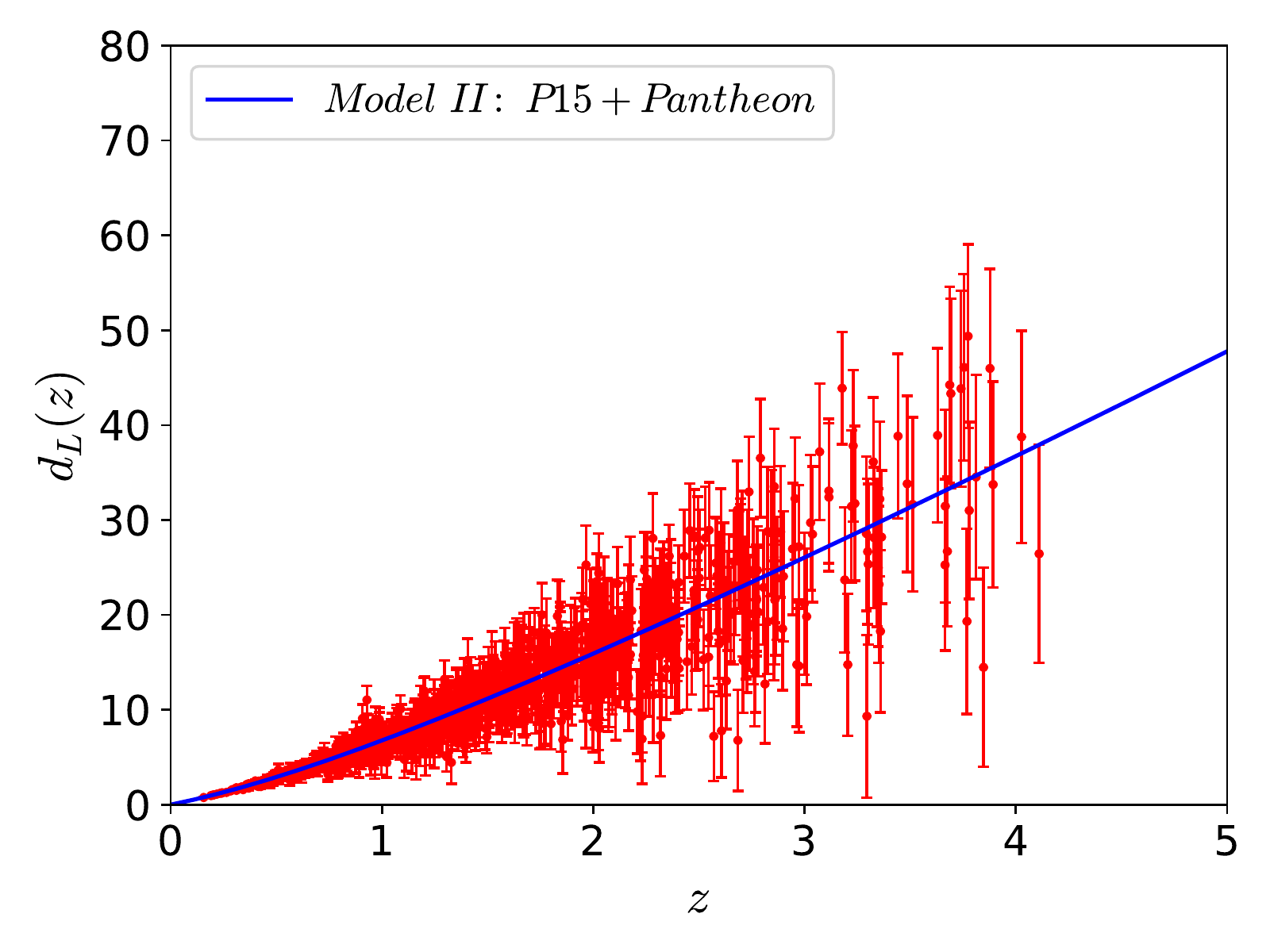}
\includegraphics[width=0.38\textwidth]{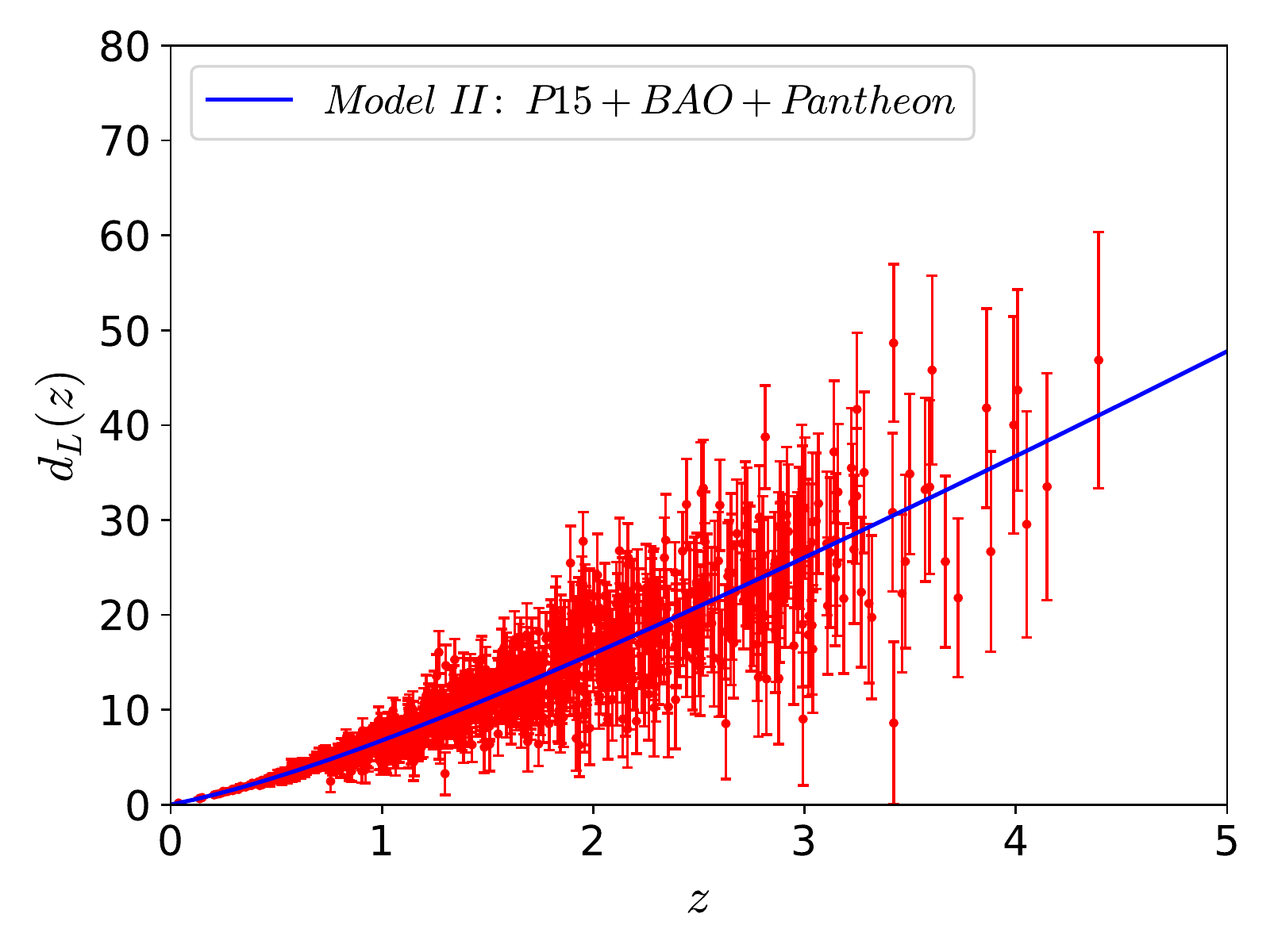}
\caption{For the fiducial (Model II) model we first constrain the cosmological parameters using the datasets P15, P15+BAO, P15+Pantheon and P15+BAO+Pantheon and then we use the best-fit of the parameters for ``each dataset'' to generate the corresponding GW catalogue. Following this, in each panel we show $d_L (z)$ vs $z$ catalogue with the corresponding error bars for 1000 simulated GW events. The upper left and upper right panels respectively present the catalogue ($z$, $d_L (z)$) with the corresponding error bars for 1000 simulated events derived using the P15 alone and P15+BAO dataset. The lower left and lower right panels respectively present the catalogue ($z$, $d_L (z)$) with the corresponding error bars for 1000 simulated events derived using the CMB+Pantheon and P15+BAO+Pantheon datasets. }
\label{dl-modelII}
\end{figure*}

Before we present all the extracted cosmological constraints from the present cosmological scenarios, we wish to present the following. To generate GW catalogue, or mock GW data, it is essential to consider the fiducial model. Here, we consider Model I and Model II as the fiducial models. Now,  
assuming Model I as the fiducial model, we first constrain it using various observational data summarized in Table \ref{tab:results1}. Then considering the 
best-fit values of  all the free and derived parameters of this model (Model I), and following \cite{Du:2018tia,Yang:2019bpr}, we generate the corresponding GW catalogue containing 1000 simulated GW events. 
In Fig.  \ref{dl-modelI} we show $d_L (z)$ vs. $z$ (with error bars on $d_L (z)$) for simulated 1000 GW events. We use this catalogue as the forecasted dataset and include them with the standard cosmological datasets, namely, P15, BAO, Pantheon for the next step of the analysis. 
In a similar fashion we generate simulated 1000 GW events for the second model in this work and Fig. \ref{dl-modelII} shows the corresponding $d_L (z)$ vs. $z$  graphics. 
In what follows we describe the observational constraints on each model considering the usual cosmological probes and the inclusion of the simulated GWSS.   
\begin{figure*}
    \centering
    \includegraphics[width=0.88\textwidth]{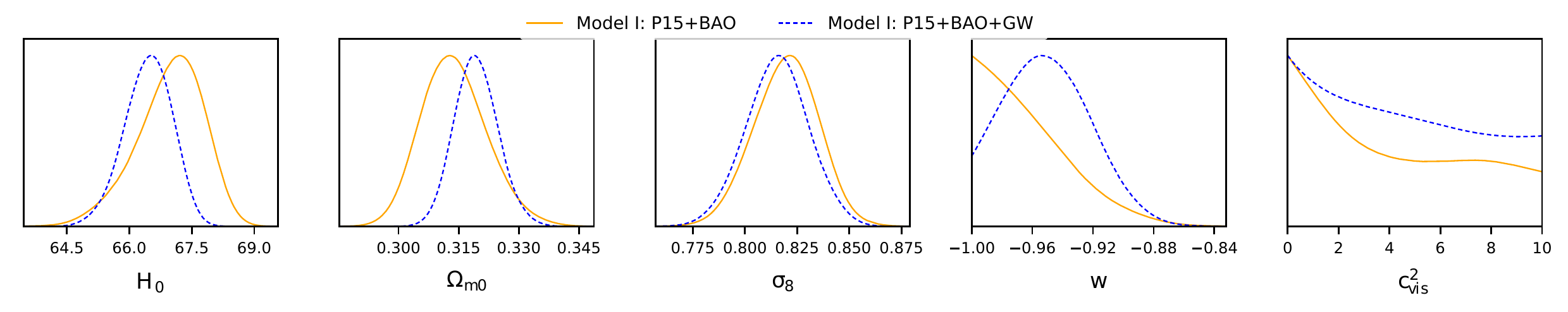}
    \includegraphics[width=0.88\textwidth]{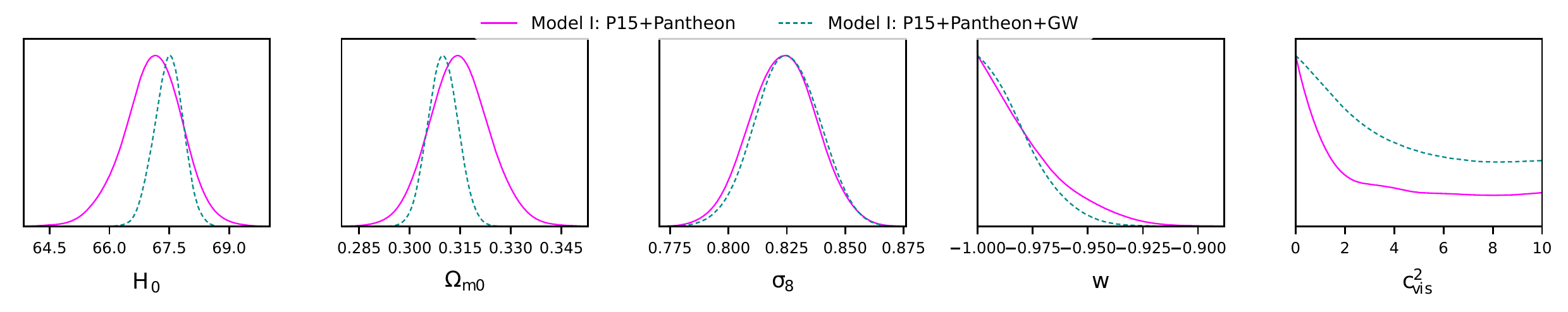}
    \includegraphics[width=0.88\textwidth]{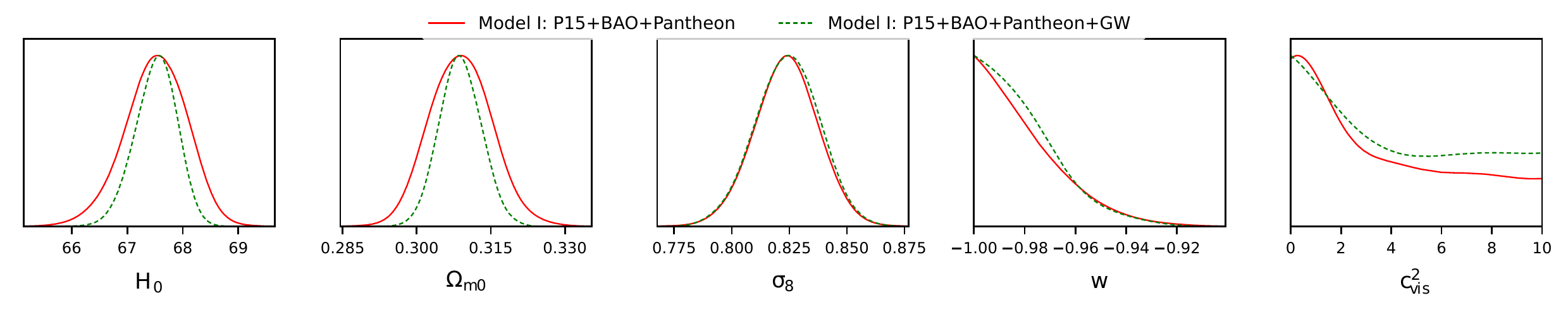}
    \caption{1-dimensional marginalized posterior distributions of some key parameters of Model I for the datasets P15+BAO,
    P15+BAO+GW (upper panel), P15+Pantheon, P15+Pantheon+GW (middle panel) and P15+BAO+Pantheon, 
    P15+BAO+Pantheon+GW (lower panel). }
    \label{fig:1D-ModelI-A}
\end{figure*}
\begin{figure}
\includegraphics[width=0.42\textwidth]{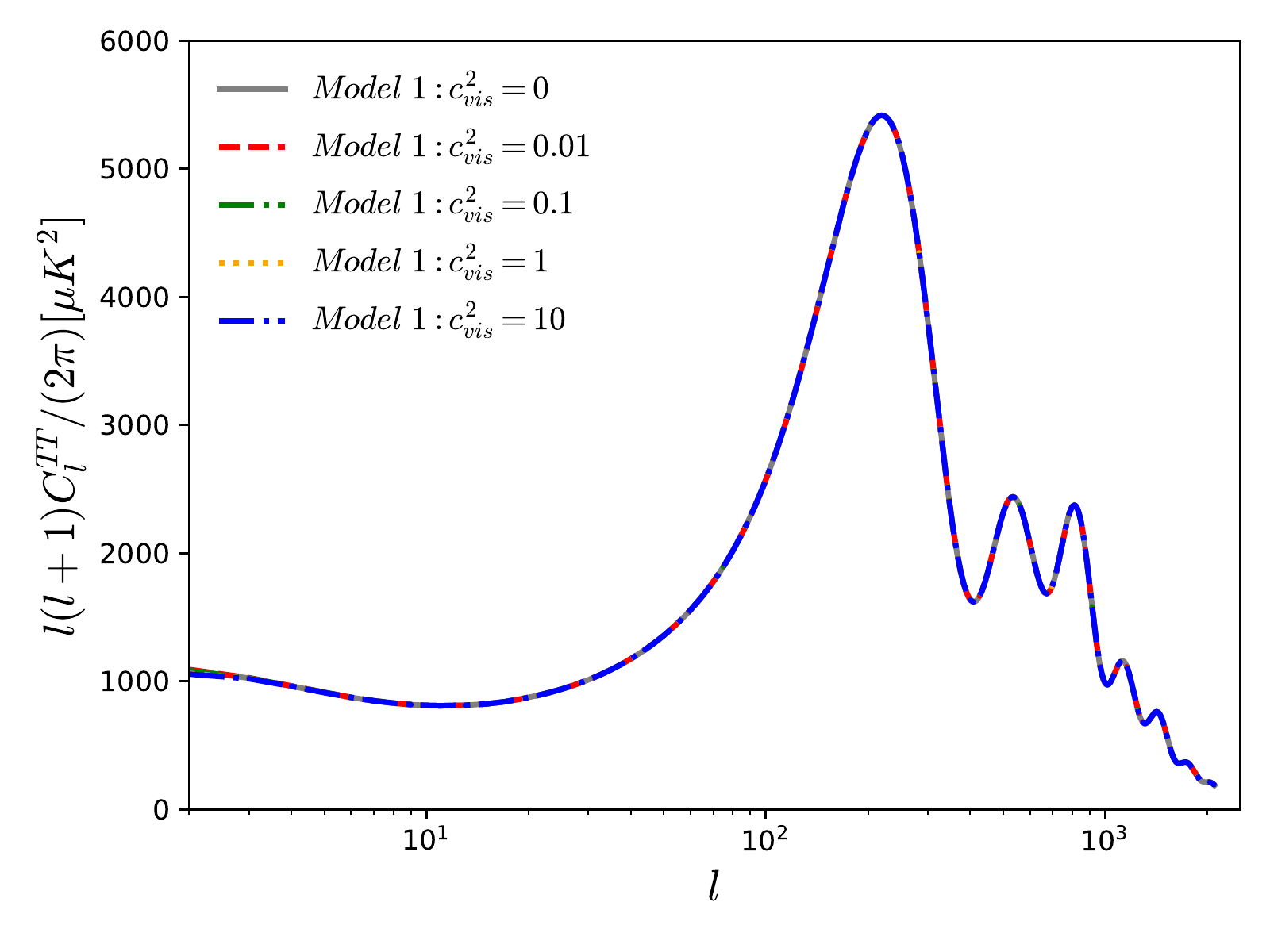}
\includegraphics[width=0.42\textwidth]{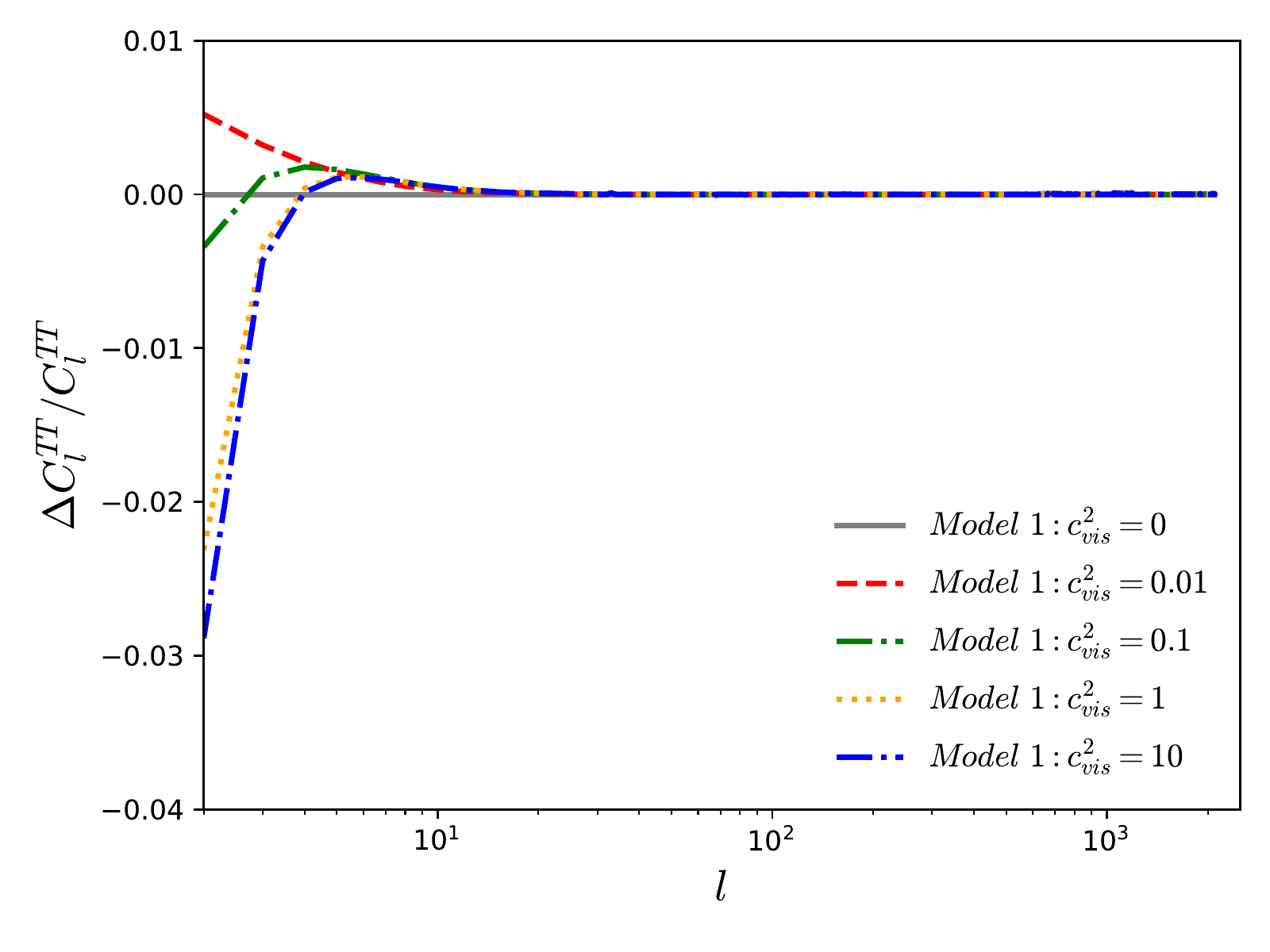}
\caption{The plots reflect the effects on the CMB TT spectra for Model I taking various values of the viscous sound speed, $c_{vis}^2$. As Model I restricts the dark energy equation of state, $w$ in the quintessence region, so we fix $w = -0.95$ and the remaining parameters have been taken  from the combined analysis P15+BAO+Pantheon (see the last column of Table \ref{tab:results1}). Note that the qualitative behavior of the plots do not change if one varies the dark energy equation of state.  From the right plot one can see that as long as $c_{vis}^2$ increases, a very mild deviation of the curves with $c_{vis}^2 \neq  0$  from the curve represening $c_{vis}^2 = 0$ (no anisotropic stress) appears. }
\label{fig-cmb-modelI}
\end{figure}
\begin{figure*}
    \centering
    \includegraphics[width=0.99\textwidth]{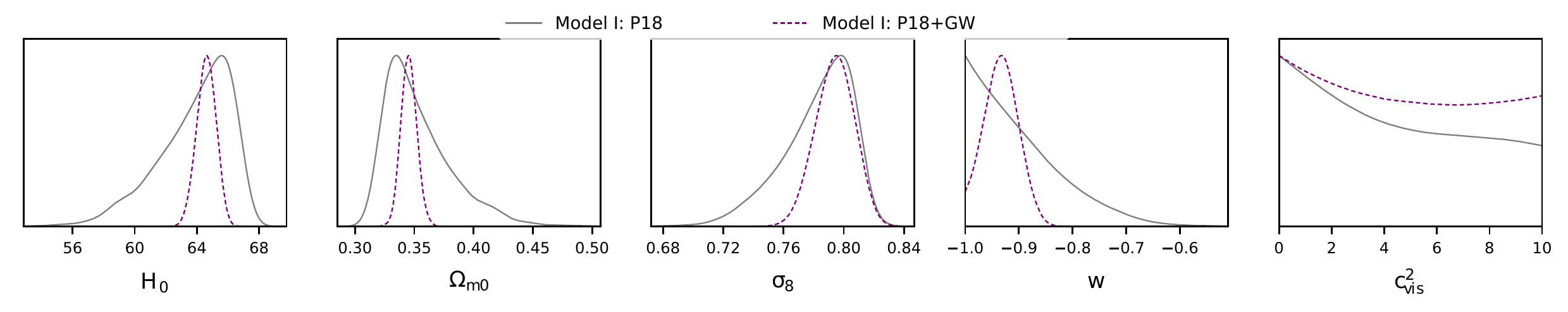}
    \caption{1-dimensional marginalized posterior distributions of some key parameters of Model I for the datasets P18 and
    P18+GW. }
    \label{fig:1D-ModelI-Planck2018}
\end{figure*}
\begin{figure*}
    \centering
    \includegraphics[width=0.99\textwidth]{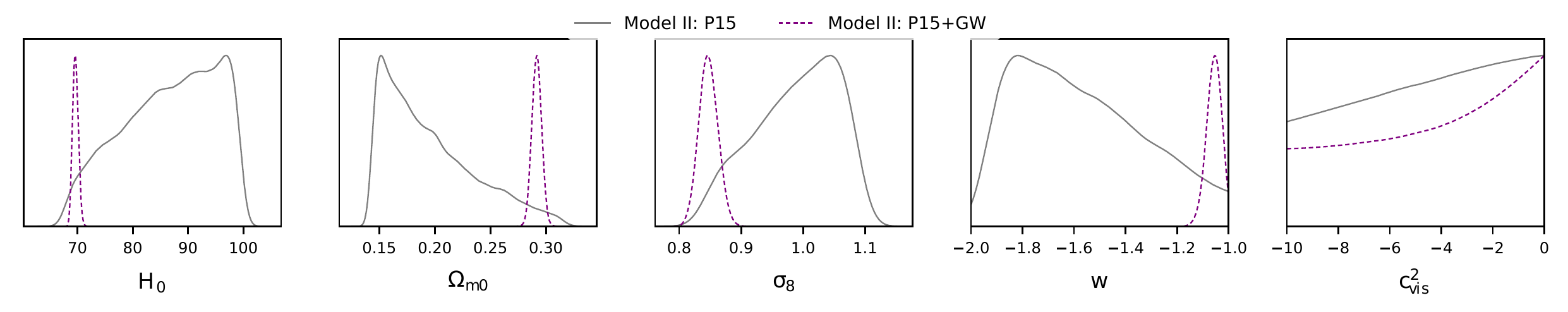}
    \caption{1-dimensional marginalized posterior distributions of some key parameters of Model II for the datasets P15 and P15+GW. }
    \label{fig:1D-ModelII}
\end{figure*}
\begin{figure*}
    \centering
    \includegraphics[width=0.38\textwidth]{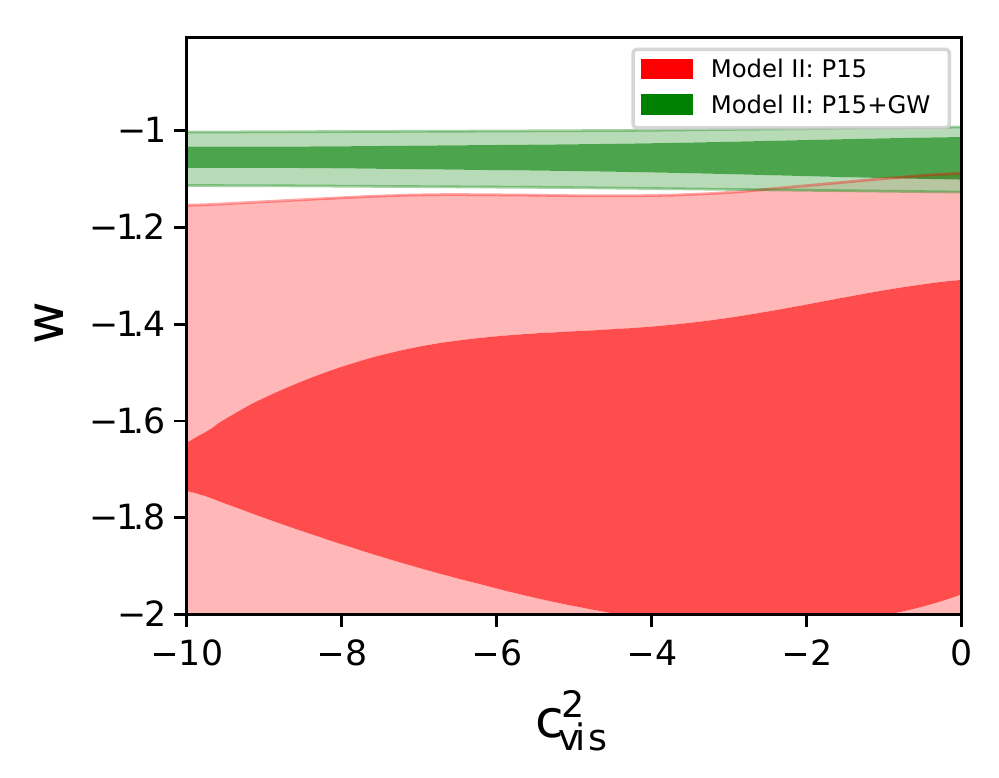}
    \includegraphics[width=0.38\textwidth]{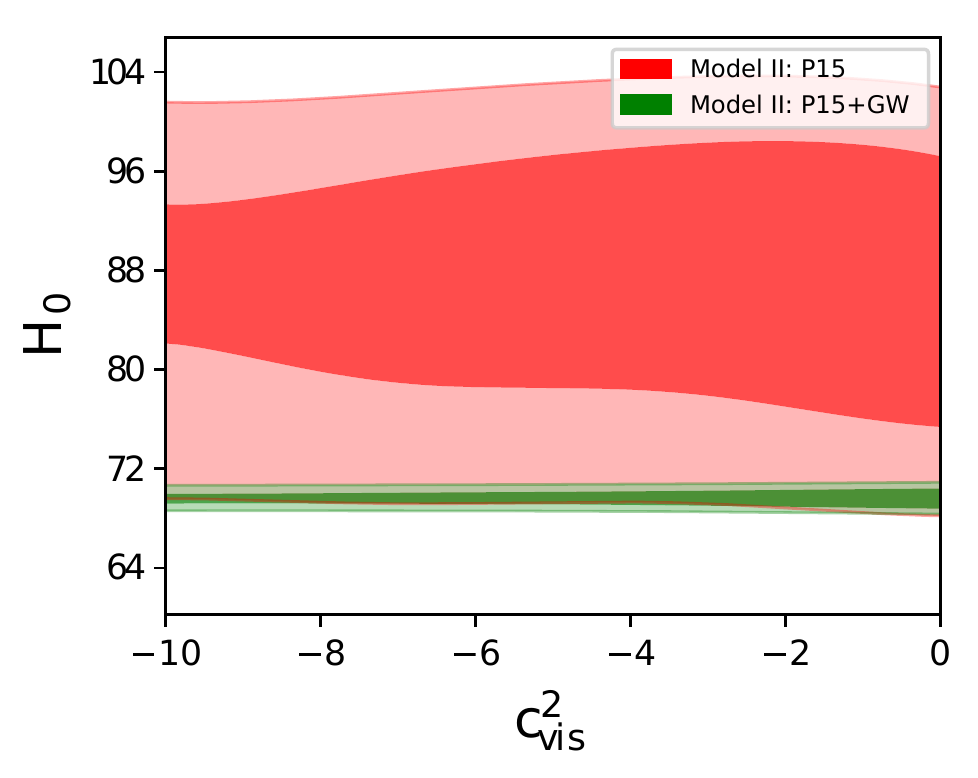}
    \includegraphics[width=0.38\textwidth]{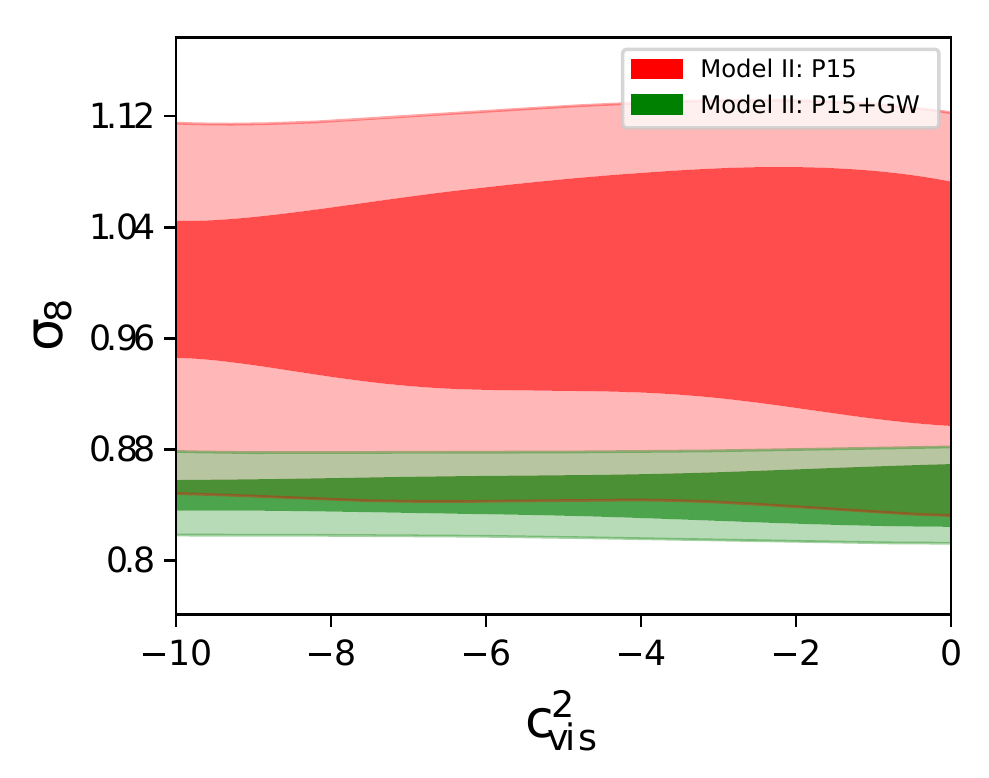}
    \includegraphics[width=0.38\textwidth]{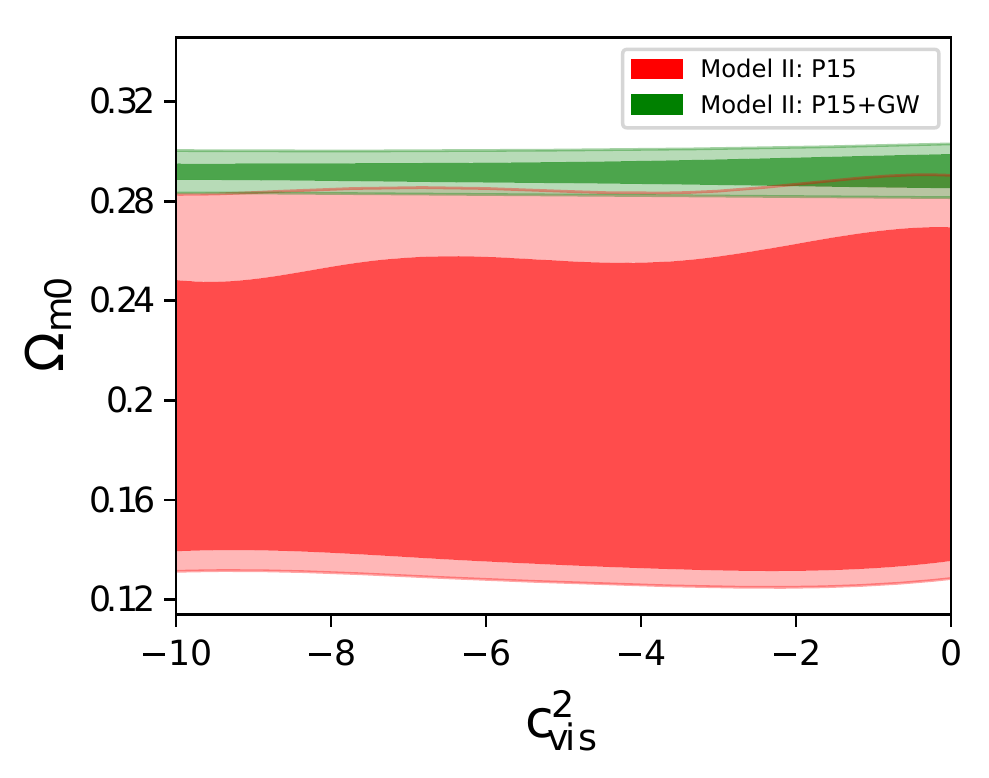}
    \caption{2-dimensional contour plots for Model II showing the dependence of $c_{vis}^2$ with other cosmological parameters for the P15 (red contours) and P15+GW (green contours) datasets. }
    \label{fig:2D-ModelII}
\end{figure*}
\begin{figure*}
    \centering
    \includegraphics[width=0.88\textwidth]{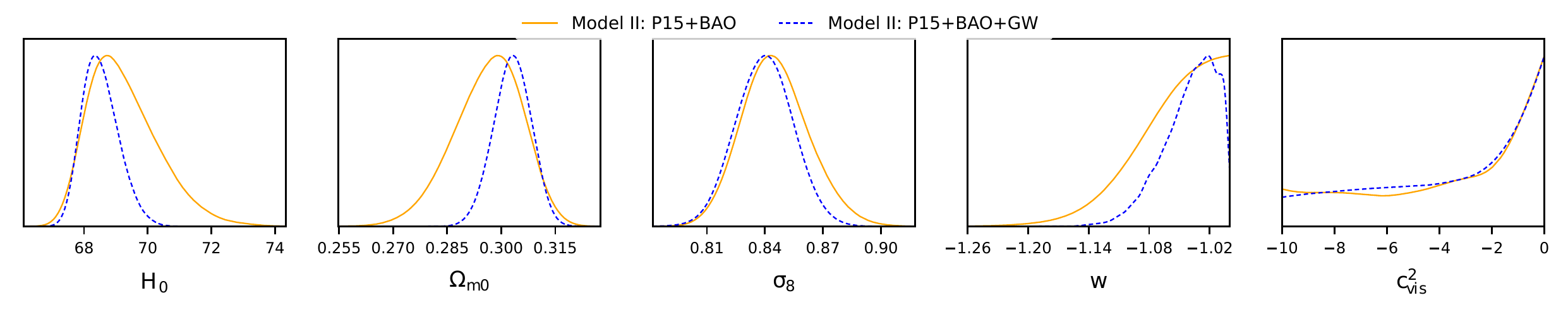}
    \includegraphics[width=0.88\textwidth]{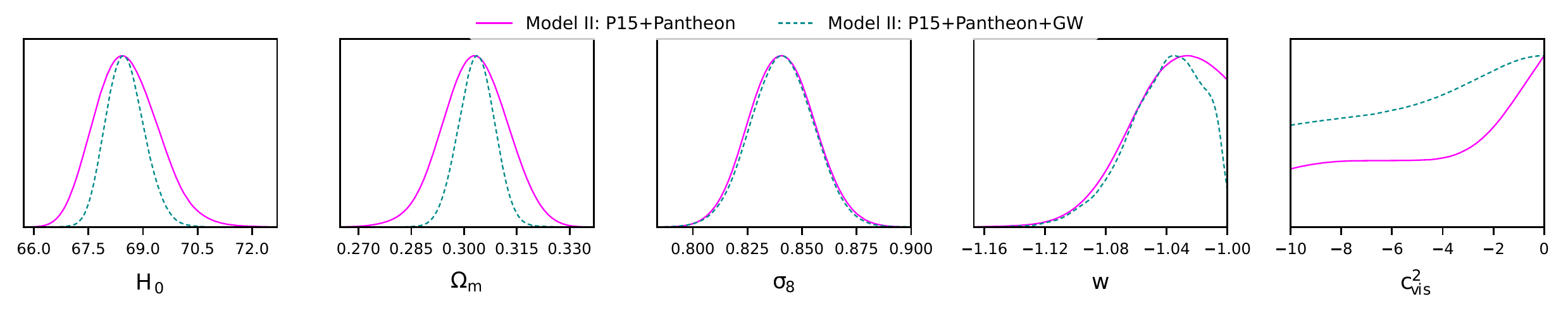}
    \includegraphics[width=0.88\textwidth]{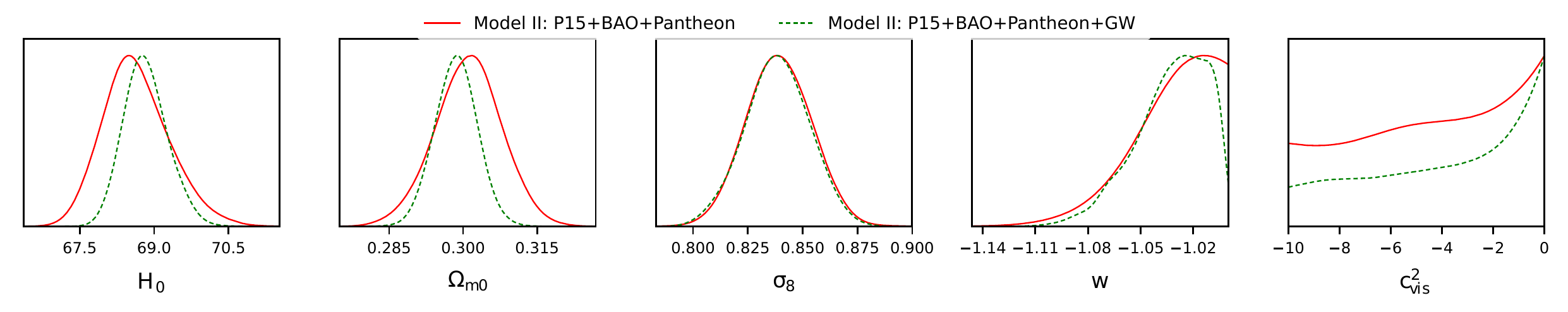}
    \caption{1-dimensional marginalized posterior distributions for some key parameters of Model II for the datasets P15+BAO,
    P15+BAO+GW (upper panel), P15+Pantheon, P15+Pantheon+GW (middle panel) and P15+BAO+Pantheon, P15+BAO+Pantheon+GW (lower panel). }
    \label{fig:1D-ModelII-A}
\end{figure*}
\begin{figure}
\includegraphics[width=0.42\textwidth]{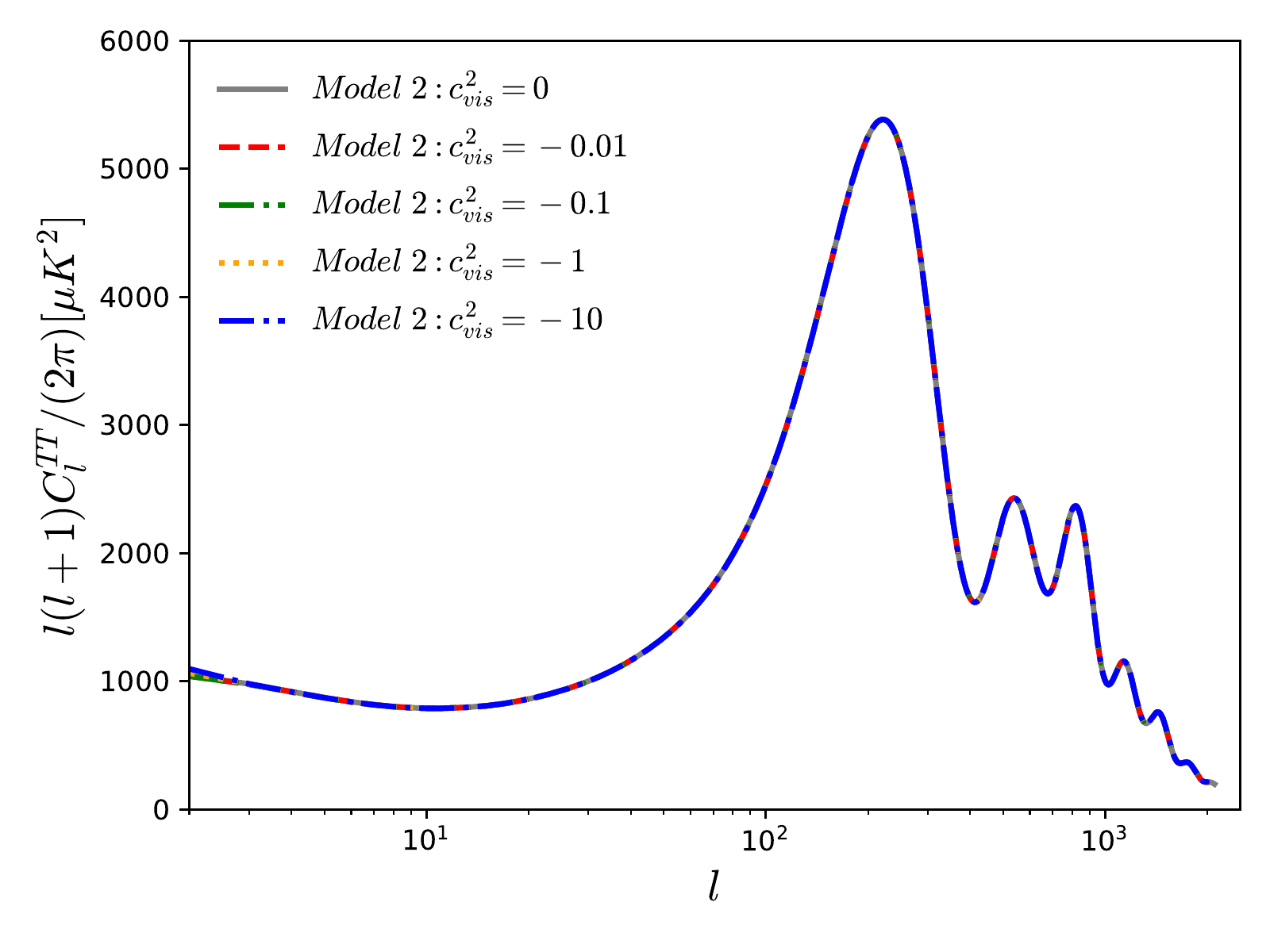}
\includegraphics[width=0.42\textwidth]{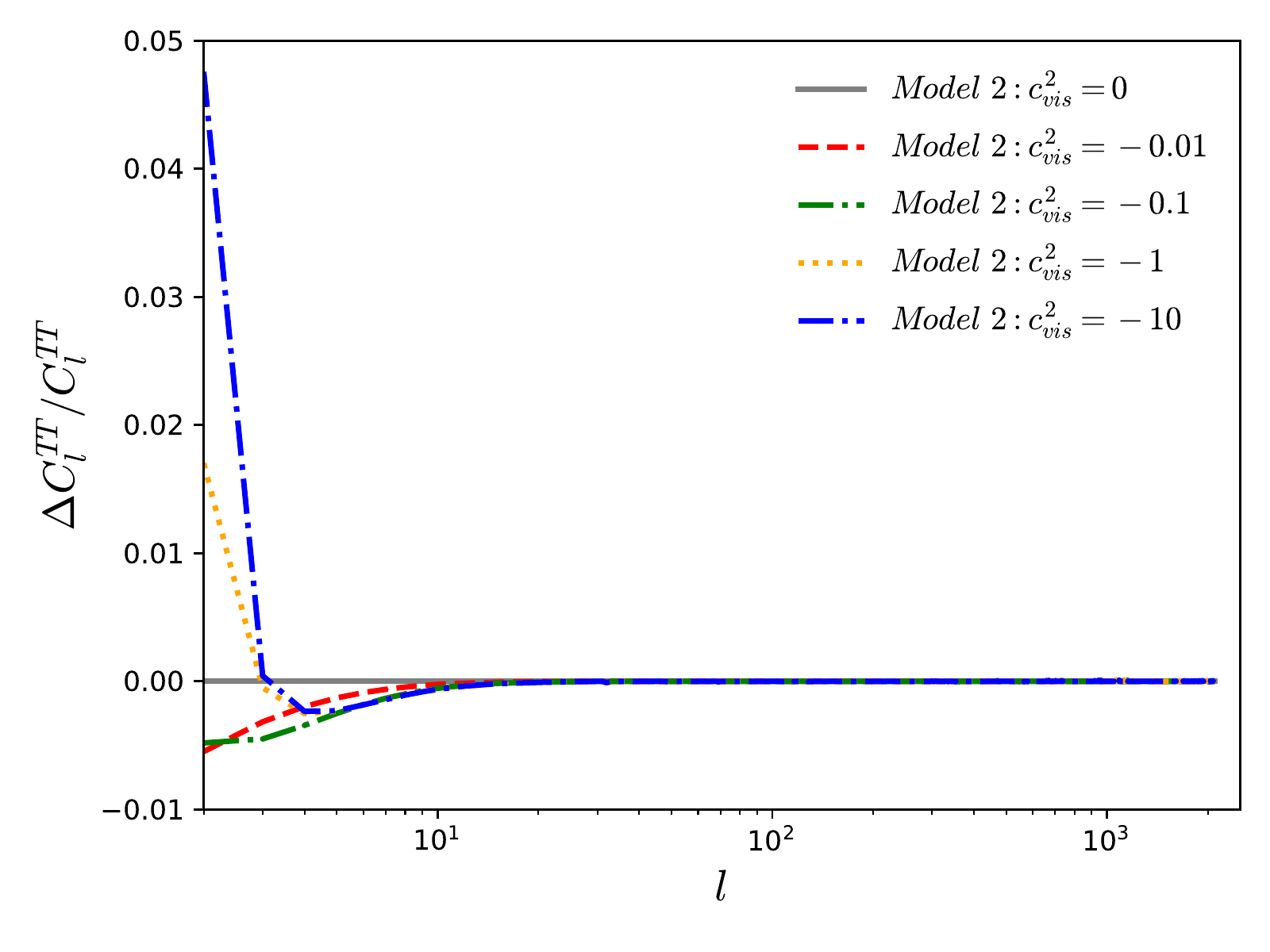}
\caption{The plots reflect the effects on the CMB TT spectra for Model II taking various values of the viscous sound speed, $c_{vis}^2$. As Model II restricts the dark energy equation of state, $w$ in the phantom region, so we fix $w = -1.1$ and the remaining parameters have been taken  from the combined analysis P15+BAO+Pantheon (see the last column of Table \ref{tab:results2}). Note that the qualitative behavior of the plots do not change if one varies the dark energy equation of state.  From the right plot one can see that as long as $c_{vis}^2$ increases, a very mild deviation of the curves with $c_{vis}^2 \neq  0$  from the curve represening $c_{vis}^2 = 0$ (no anisotropic stress) appears.  }
\label{fig-cmb-modelII}
\end{figure}
\begin{figure*}
    \centering
    \includegraphics[width=0.99\textwidth]{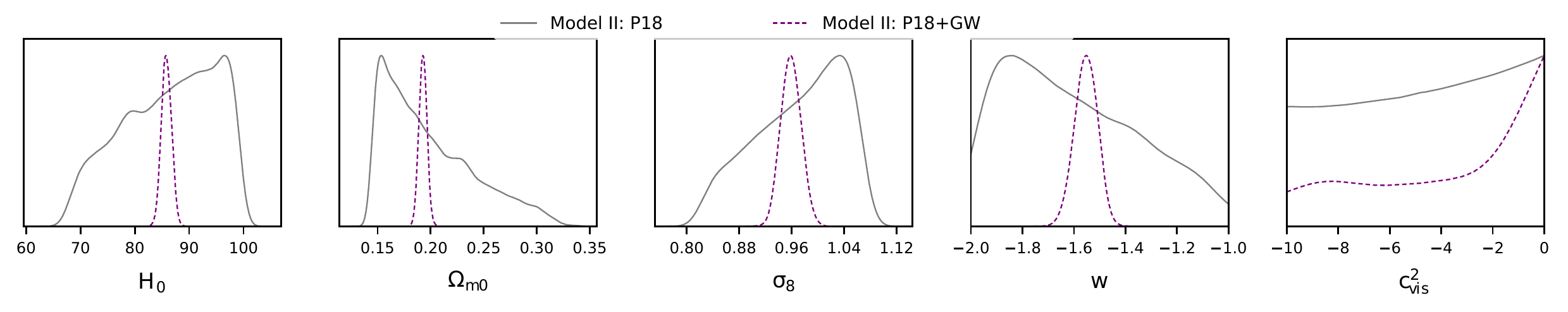}
    \caption{1-dimensional marginalized posterior distributions for some parameters of Model II for the datasets P18 and P18+GW. }
    \label{fig:1D-ModelII-Planck2018}
\end{figure*}
\begingroup                                                                                                                     
%\squeezetable                                                                                                                   
\begin{center}                                                                                                                  
\begin{table*}
\scalebox{0.9}
{                                                                                                                                                                                                                                      
\begin{tabular}{cccccccccccccccccc}                                                                                                            
\hline\hline                                                                                                                    
Parameters & P15 &  P15+BAO & P15+Pantheon & P15+BAO+Pantheon \\ \hline
$\Omega_c h^2$ & $    0.1191_{-    0.0014-    0.0027}^{+    0.0014+    0.0027}$ & $    0.1191_{-    0.0011-    0.0021}^{+    0.0011+    0.0022}$ & $    0.1195_{-    0.0013-    0.0026}^{+    0.0014+    0.0026}$ & $    0.1189_{-    0.0011-    0.0022}^{+    0.0010+    0.0022}$ \\

$\Omega_b h^2$ & $    0.02228_{-    0.00016-    0.00030}^{+    0.00016+    0.00031}$ & $    0.02226_{-    0.00015-    0.00028}^{+    0.00014+    0.00029}$ & $    0.02224_{-    0.00014-    0.00029}^{+    0.00014+    0.00029}$ & $    0.02228_{-    0.00015-    0.00027}^{+    0.00014+    0.00029}$ \\

$100\theta_{MC}$ & $    1.04080_{-    0.00033-    0.00063}^{+    0.00033+    0.00062}$ & $    1.04080_{-    0.00031-    0.00059}^{+    0.00030+    0.00062}$ & $    1.04073_{-    0.00032-    0.00062}^{+    0.00031+    0.00064}$ & $    1.04083_{-    0.00031-    0.00060}^{+    0.00031+    0.00062}$  \\

$\tau$ & $    0.076_{-    0.017-    0.034}^{+    0.018+    0.034}$ & $    0.080_{-    0.017-    0.032}^{+    0.017+    0.033}$ & $    0.078_{-    0.017-    0.032}^{+    0.017+    0.033}$ & $    0.082_{-    0.017-    0.032}^{+    0.017+    0.033}$ \\

$n_s$ & $    0.9665_{-    0.0046-    0.0089}^{+    0.0046+    0.0093}$ & $    0.9665_{-    0.0040-    0.0079}^{+    0.0040+    0.0077}$ & $    0.9655_{-    0.0044-    0.0086}^{+    0.0044+    0.0087}$ & $    0.9671_{-    0.0039-    0.0078}^{+    0.0042+    0.0077}$  \\

${\rm{ln}}(10^{10} A_s)$ & $    3.085_{-    0.034-    0.066}^{+    0.034+    0.066}$ & $    3.093_{-    0.033-    0.062}^{+    0.033+    0.064}$ & $    3.091_{-    0.034-    0.064}^{+    0.033+    0.065}$ & $    3.095_{-    0.032-    0.065}^{+    0.033+    0.063}$ \\

$w$ & $ > -1.917 > -1.973$ & $ > -1.071 > -1.133$ & $ > -1.052 > -1.088 $  & $ > -1.041 > -1.078 $ \\

$c^2_{vis}$ & $ \mbox{unconstrained}$ &  $ \mbox{unconstrained}$ & $ \mbox{unconstrained}$ & $ \mbox{unconstrained}$ \\

$\Omega_{m0}$ & $    0.195_{-    0.055-    0.063}^{+    0.020+    0.088}$ & $    0.296_{-    0.008-    0.020}^{+    0.011+    0.018}$ & $    0.303_{-    0.009-    0.018}^{+    0.009+    0.018}$ & $    0.301_{-    0.007-    0.014}^{+    0.007+    0.013}$  \\

$\sigma_8$ & $    0.989_{-    0.052-    0.139}^{+    0.094+    0.116}$ &  $    0.845_{-    0.019-    0.034}^{+    0.016+    0.036}$ & $    0.841_{-    0.015-    0.029}^{+    0.015+    0.030}$ &  $    0.839_{-    0.015-    0.029}^{+    0.015+    0.029}$ \\

$H_0$ & $   86.69_{-    6.03-   16.53}^{+   12.21+   14.14}$ & $   69.25_{-    1.38-    2.01}^{+    0.80+    2.33}$ &  $   68.57_{- 0.97- 1.70}^{+ 0.80+    1.73}$ & $   68.64_{-    0.76-    1.32}^{+ 0.60+    1.40}$ \\

\hline\hline                                                                                                                    
\end{tabular}
}                                                                                                                   
\caption{The table presents the constraints on various free and derived cosmological parameters at 68\% and 95\% CL for Model II using the usual cosmological probes, namely, 
P15, BAO and Pantheon. For the dark energy equation of state we present its upper limits at 68\% and 95\% CL. }
\label{tab:results2}                                                                                                   
\end{table*}                                                                                                                     
\end{center}                                                                                                                    
\endgroup 
\begingroup                                                                                                                     
%\squeezetable                                                                                                                   
\begin{center}                                                                                                                  
\begin{table*}
\scalebox{0.9}
{                                                                                                                                                                                                                                      
\begin{tabular}{cccccccccccccccc}                                                                                                            
\hline\hline                                                                                                                    
Parameters & P15+GW & P15+BAO+GW & P15+Pantheon+GW & P15+BAO+Pantheon+GW\\ \hline

$\Omega_c h^2$ & $    0.1184_{-    0.0012-    0.0023}^{+    0.0012+    0.0024}$ & $    0.1194_{-    0.0011-    0.0019}^{+    0.0010+    0.0020}$ &  $    0.1196_{-    0.0012-    0.0021}^{+    0.0011+    0.0023}$ & $    0.1184_{-    0.0010-    0.0020}^{+    0.0009+    0.0020}$ \\

$\Omega_b h^2$ & $    0.02235_{-    0.00014-    0.00027}^{+    0.00014+    0.00027}$ & $    0.02220_{-    0.00013-    0.00026}^{+    0.00013+    0.00026}$ & $    0.02222_{- 0.00013-    0.00025}^{+    0.00013+    0.00025}$ & $    0.02233_{-    0.00013-    0.00026}^{+    0.00013+    0.00025}$ \\

$100\theta_{MC}$ & $    1.04091_{-    0.00030-    0.00061}^{+    0.00030+    0.00059}$ & $    1.04069_{-    0.00031-    0.00061}^{+    0.00030+    0.00059}$ &  $    1.04074_{-    0.00030-    0.00059}^{+    0.00030+    0.00057}$ & $    1.04089_{-    0.00029-    0.00057}^{+    0.00029+    0.00058}$ \\

$\tau$ & $    0.085_{-    0.017-    0.033}^{+    0.017+    0.033}$ & $    0.078_{-    0.016-    0.033}^{+    0.016+    0.031}$ & $    0.078_{-    0.016-    0.033}^{+    0.016+    0.032}$ & $    0.083_{-    0.016-    0.033}^{+    0.016+    0.031}$ \\

$n_s$ & $    0.9683_{-    0.0042-    0.0085}^{+    0.0043+    0.0080}$ & $    0.9657_{-    0.0037-    0.0073}^{+    0.0037+    0.0073}$ & $    0.9652_{-    0.0039-    0.0078}^{+    0.0042+    0.0073}$ & $ 0.9683_{-    0.0038-    0.0079}^{+    0.0038+    0.0073}$  \\

${\rm{ln}}(10^{10} A_s)$ & $    3.101_{-    0.033-    0.065}^{+    0.033+    0.064}$ & $    3.089_{-    0.031-    0.064}^{+    0.031+    0.061}$ & $    3.090_{-    0.033-    0.064}^{+    0.032+    0.062}$ & $    3.098_{-    0.032-    0.066}^{+    0.035+    0.066}$ \\

$w$ & $ > -1.084 > -1.114$ & $ > -1.055 > -1.093 $ & $ > -1.06 > -1.091 $ & $ > -1.046 > -1.075$  \\

$c^2_{vis}$ & $\mbox{unconstrained}$ & $   \mbox{unconstrained}$ & $\mbox{unconstrained}$ & $\mbox{unconstrained}$ \\

$\Omega_{m0}$ & $    0.292_{-    0.004-    0.008}^{+    0.004+    0.008}$ &  $    0.303_{-    0.005-    0.010}^{+    0.006+    0.010}$ & $    0.304_{-    0.005-    0.010}^{+    0.005+    0.010}$ &  $    0.299_{-    0.004-    0.008}^{+    0.004+    0.008}$ \\

$\sigma_8$ & $    0.847_{-    0.016-    0.030}^{+    0.015+    0.031}$ & $    0.840_{-    0.015-    0.029}^{+    0.015+    0.029}$ & $    0.841_{-    0.014-    0.028}^{+    0.014+    0.028}$ & $    0.838_{-    0.015-    0.030}^{+    0.015+    0.029}$ \\

$H_0$ & $   69.63_{-    0.57-    0.94}^{+    0.48+    1.05}$ & $   68.52_{-    0.67-    1.08}^{+    0.48+    1.18}$ & $   68.52_{-    0.59-    1.04}^{+    0.49+    1.09}$ & $   68.83_{-    0.50-    0.82}^{+    0.41+    0.93}$ \\
\hline\hline                                                                                                                    
\end{tabular}
}                                                                                                                   
\caption{In this table we show the constraints on various free and derived cosmological parameters of Model II at 68\% and 95\% CL after the inclusion of GWSS data with the standard cosmological probes P15, BAO and Pantheon.}
\label{tab:results2a}                                                                                                   
\end{table*}                                                                                                                     
\end{center}                                                                                                                    
\endgroup

\subsection{Model I: $c_{vis}^2> 0$, $w> -1$}
\label{subsec-model1}

In Table \ref{tab:results1} we show the constraints on the model parameters for the usual cosmological probes and in Table \ref{tab:results1a} we display the constraints on the model parameters after the inclusion of the simulated GWs data with the usual cosmological probes. Thus, Table \ref{tab:results1} and Table \ref{tab:results1a} summarize the main results on this model.  In the following we become more explicit on the improvements of the constraints, if any, after the inclusion of GWs to the usual cosmological probes mentioned above.

Let us first focus on the constraints from P15 and P15+GW. From Table \ref{tab:results1} we notice see that for P15 data alone the Hubble constant at present, i.e. $H_0$ takes lower value compared to $\Lambda$CDM based Planck but with high error bars  compared to what we find in $\Lambda$CDM based Planck \cite{Ade:2015xua}. In particular, one finds that for P15 alone $H_0 = 64.11_{- 1.70}^{+    3.22} $ (68\% CL). When the simulated GW data are added to P15, the Hubble constant rises up giving $H_0 = 66.94_{-    0.38}^{+    0.39}$ (68\% CL, P15+GW). One can clearly see that the inclusion of GW to P15 significantly reduces the error bars on $H_0$. In fact, the error bars on $H_0$ are reduced at least by a factor of $5$. This actually reflects the constraining power of GW.  In a similar fashion when we look at the other derived parameters of this model, namely,  $\Omega_{m0}, \sigma_8$, one can draw similar conclusion, that means the effects of GWs on the cosmological parameters is transparent.  In fact, the free parameter, $w$, is also affected significantly after the addition of GW to P15.  We see that the 68\% upper bound on $w$  for P15 alone is $w < -0.854$, which is significantly changed to $w< -0.974$ after the addition of GW to P15. Now, concerning the viscous speed of sound, $c_{vis}^2$, we find that this parameter is neither constrained by P15 alone nor the addition of GW to P15 helps to constrain it. We refer to Fig. \ref{fig:1D-ModelI} showing the 1D marginalized posterior distributions for some key parameters of this scenario for P15 and P15+GW datasets.  
One may wonder that perhaps the increase of the prior on $c_{vis}^2$ may help to constrain, however, this is not true in this case. We found that even if the prior varies in the interval $[0, 100]$, 
this parameter remains unconstrained. We also  
show Fig. \ref{fig:2D-ModelI} displaying the dependence of $c_{vis}^2 $ with other parameters for P15 and P15+GW datasets. Thus, we find that even if we add mock GW data to CMB from P15, GW data do not add any extra constraining power to CMB from P15 which might constrain $c_{vis}^2$.

We now discuss the next two datasets, namely P15+BAO and its companion P15+BAO+GW. For a quick view on the cosmological constraints we refer to the third columns of both Table \ref{tab:results1} and Table \ref{tab:results1a}. Additionally, for graphical views, we refer to the upper panel of Fig. \ref{fig:1D-ModelI-A} which shows the 1D marginalized posterior distributions for some selected parameters. We find that the addition of GW to P15+BAO shifts the higest peak of $H_0$ towards higher values and shifts $\Omega_{m0}$ towards lower values. This is consistent since there is already a known correlation between $H_0$ and $\Omega_{m0}$. So, the addition of GW does not alter such correlation.  A similar but very small shift of the $\sigma_8$ parameter is also observed. Concerning the dark energy equation of state, $w$, we have an interesting observation. From the 1D posterior distribution of $w$ (upper panel of Fig. \ref{fig:1D-ModelI-A}) we see that after the inclusion of GW to P15+BAO, we find the highest peak of $w$ which was absent for the usual CMB+BAO analysis. Finally, we notice that the parameter $c_{vis}^2$ remains unconstrained for both the datasets, namely, 
P15+BAO and P15+BAO+GW. So, we see that the addition of GW to CMB+BAO does not alter the nature of this parameter.

We now consider the following two cases, namely CMB+Pantheon and CMB+Pantheon+GW. The results are summarized in the fourth columns of both Table \ref{tab:results1} and Table \ref{tab:results1a}. And we refer to the middle panel of Fig. \ref{fig:1D-ModelI-A} for  a quick look on the 1D posterior distributions of some important parameter before and after the inclusion of GW to the corresponding dataset (i.e., P15+Patheon).  
Our results are  very clear and straightforward.  In a similar fashion to the previous two analyses (i.e. P15+BAO and P15+BAO+GW), here too, we find that the addition of GW to P15+Pantheon shifts the highest peak of the Hubble constant towards higher values having an additional shift of $\Omega_{m0}$ towards its  lower values. However, the parameter space of  both $H_0$ and $\Omega_{m0}$ are certainly improved due to GW. In addition we do not find any changes to the parameter space of $\sigma_8$, which is clear if one looks at the 1D posterior distribution of this parameter. Moreover, we have a different result when one looks at the 1D posterior of $w$ for both 
CMB+Pantheon and P15+Pantheon+GW.  One could see that in contrary to the previous observation with P15+BAO+GW, the peak of $w$ disappears in this case. Finally, our conclusion regarding the viscous sound speed, $c_{vis}^2$ remains same, that means it is again unconstrained for both the datasets. 

Lastly, we come to the last two datasets in this series, namely, P15+BAO+Pantheon and P15+BAO+Pantheon+GW. The last columns of both Table \ref{tab:results1} and Table \ref{tab:results1a} summarize the constraints on the model parameters and the lower panel of Fig. \ref{fig:1D-ModelI-A} displays the 1D posterior distributions of some important parameters of this model.  Looking at the lower panel of Fig. \ref{fig:1D-ModelI-A}, specially for $H_0$ and $\Omega_{m0}$ we find their improvements after the 
inclusion of GW, however, we do not find any shifts of the highest peaks of $H_0$ and $\Omega_{m0}$ in their posterior distributions in contrary to the earlier cases, such as P15+BAO+GW and P15+Pantheon+GW. Here we again see that the parameter $c_{vis}^2$ is still unconstrained for both the datasets. So, GW data seem to be unable to constrain this particular parameter.  One can try to understand this unconstrained nature of the viscous sound speed through Fig. \ref{fig-cmb-modelI} where we have displayed how various values of $c_{vis}^2$ can affect the temperature anisotropy in the CMB spectra. The left plot of Fig. \ref{fig-cmb-modelI} shows the CMB TT spectra while in the right plot of Fig. \ref{fig-cmb-modelI} we have shown the corresponding residual plot.  As one can see, in the lower multipoles region ($l < 10$) a very mild deviation in the CMB TT spectra from $c_{vis}^2  = 0$ appears (see the right plot of Fig. \ref{fig-cmb-modelI}) due to the Integrated Sachs-Wolfe (ISW) effects coming from non-zero anisotropic 
stress in DE, see also  \cite{Chang:2014mta}.

When we finished all the analyses of this paper, Planck released its final CMB data \cite{Aghanim:2018oex,Aghanim:2019ame}. We then wanted to check  whether the new CMB data from Planck 2018 final release (P18 as referred in the text) could constrain the parameter, $c_{vis}^2$ of this scenario.  We found that P18 data are also unable to constrain $c_{vis}^2$ which is also unconstrained by the earlier P15 data.
To illustrate this nature, in Fig. \ref{fig:1D-ModelI-Planck2018} we have shown the 1-dimensional posterior distributions of some key parameters of this model showing the constraining power of P18 data alone and also the effects of GW  on $c_{vis}^2$ after its inclusion to P18.  From the present analyses, we have clearly visualized that the unconstrained nature of $c_{vis}^2$ is not controlled by any of the employed external datasets, such as BAO, Pantheon, etc. And we have seen that when P15 data are unable to constrain $c_{vis}^2$, this parameter remains unconstrained by the external datasets. Thus, since P18 data remain unable to constrain $c_{vis}^2$, there is no reason to consider other external datasets with P18 in order to check whether the viscous sound speed will be constrained or not.  Hence, we do not consider these combinations for this work. The nature of $c_{vis}^2$ will actually be the same in this case.

\subsection{Model II: $c_{vis}^2 < 0$, $w< -1$}
\label{susbsec-model2}

In Table \ref{tab:results2} we show the constraints on the model parameters using the usual cosmological probes and in Table \ref{tab:results2a} we show the constraints on the model parameters after the inclusion of the simulated GWs data to the usual cosmological probes. Thus, Table \ref{tab:results2} and Table \ref{tab:results2a} summarize the main results on this model.

Now, following the similar pattern as with Model I we analyze this model as well. 
Thus, we first focus on the datasets namely P15 and P15+GW and discuss how the GWs data could improve the constraints on various free and derived parameters of this model. Looking at the constraint on $H_0$, one can quickly realize the effects of GWs onto it. From P15 alone, $H_0 = 86.69_{- 6.03}^{+   12.21}$ (68\% CL) while when GWs are added to P15, then $H_0$ is reduced both in its mean values as well as in its error bars, $H_0 = 69.63_{-    0.57}^{+    0.48}$ (68\% CL, P15+GW). In fact, 
the error bars on $H_0$ are reduced by a factor of more than $10$. Thus, a real effect on the $H_0$ parameter for the introduction of GWs data is clearly visible.  Similar effects on other cosmological parameters are equally evident.  As one can see that the equation-of-state for dark energy is significantly improved after the addition of GWs to P15. More explicitly, the upper limit (at 68\% CL) of the dark energy equation of state changes from $w> -1.917$ (P15) to $w>-1.084 $ (P15+GW).  Thus, looking at the constraints on $w$, one can clearly conclude that the inclusion of GWs to P15 not only decreases the mean values of $w$ taken for P15 data alone, but also this reduces the error bars of $w$ that arise from P15 data only.  Finally, we come to the most important part of this work, namely the behaviour of $c^2_{vis}$. We found that neither CMB nor P15+GW could constrain this parameter.  Exactly similar conclusion has been found for Model I. So, irrespective of the dark energy equation of state, this parameter remains constrained. In Fig. \ref{fig:1D-ModelII} we show the 1D marginalized posterior distributions of some important parameters discussed just above for the P15 and P15+GW datasets. Moreover, in order to explicitly present the behaviour of $c_{vis}^2$ with other parameters, we present the 2D joint contours in Fig. \ref{fig:2D-ModelII}.   

For rest of the analyses with other cosmological datasets, we refer to Fig. \ref{fig:1D-ModelII-A} showing the 1D marginalized posterior distributions of some selected parameters. In particular, the upper panel of Fig. \ref{fig:1D-ModelII-A} compares the constraints on the model parameters for P15+BAO and its companion P15+BAO+GW. The lower panel of Fig. \ref{fig:1D-ModelII-A} similarly compares the constraints from P15+Pantheon and P15+Pantheon+GW and finally the last panel of Fig. \ref{fig:1D-ModelII-A} compares the constraints of some selective model parameters for P15+BAO+Pantheon and P15+BAO+Pantheon+GW.  From Fig. \ref{fig:1D-ModelII-A} it is clear that due to the inclusion of GW to the standard cosmological datasets, some of the model parameters  are affected, for instance the effects on $H_0$ and $\Omega_{m0}$ are pretty clear while the effects on $w$ are not so pronounced much.  
However, the parameter on which we concentrate our focus in this work, namely, $c_{vis}^2$, is still unconstrained irrespective of either the usual cosmological probes or the inclusion of GW to them. So, effectively even if we include the GW data into the standard cosmological probes, this specific parameter remains unconstrained, that means GW fails to constrain it.  
Similar to Model I, in Fig. \ref{fig-cmb-modelII}  we show the temperature anisotropy in the CMB spectra for various values of $c_{vis}^2$. The left plot of Fig. \ref{fig-cmb-modelII} shows the CMB TT spectra whilst the right plot of Fig. \ref{fig-cmb-modelII} shows the corresponding residual plot. Similarly, we notice a very mild  deviation in the CMB TT spectra in the lower multipoles region ($l < 10$)  from $c_{vis}^2  = 0$ (see the right plot of Fig. \ref{fig-cmb-modelI}) due to the Integrated Sachs-Wolfe (ISW) effects coming from non-zero anisotropic 
stress in DE, also see \cite{Chang:2014mta}.   

Finally, similar to Model I, here too, we consider the analyses with P18 and P18+GW. The results are shown in Fig. \ref{fig:1D-ModelII-Planck2018}. One can clearly see that P18 data are unable to constrain the parameter $c_{vis}^2$ which is unconstrained also by the  P15 data.

\section{Conclusion and the Final Remarks}
\label{sec-conclu}

In this article we have considered a very general cosmological framework in which the dark energy component has an anisotropic stress. Our aim is to constrain the anisotropic stress of dark energy in terms of the viscous sound speed $c^2_{vis}$ using the usual cosmological probes, namely, CMB from Planck 2015 (P15), CMB from Planck 2018 (P18), BAO, Pantheon and then measure the constraining power of GWs data.  We have considered two different scenarios as follows: (i) Model I  where $c_{vis}^2 > 0$, and $w> -1$ and (ii) Model II  in which $c_{vis}^2 < 0$, and $w< -1$.

We started our analyses with a number of cosmological probes, such as,
P15, P15+BAO, P15+Pantheon and P15+BAO+Pantheon  and afterwards we measured the effects  of the  GW data from Einstein Telescope for the combinations P15+GW, P15+BAO+GW, P15+Pantheon+GW and P15+BAO+Pantheon+GW.  
For Model I, the results of our analyses have been shown in Table \ref{tab:results1} (without GWSS) and Table \ref{tab:results1a} (with GWSS).  In order to understand the effects of GWSS on the underlying scenario, we have shown the 1D posterior distributions of some key parameters in  Fig. \ref{fig:1D-ModelI} (only for P15 and P15+GW) and in Fig. \ref{fig:1D-ModelI-A} (for the remaining datasets). For Model II, Table \ref{tab:results2} (without GWSS),  Table \ref{tab:results2a} (with GWSS) summarize the main observational constraints on the parameters. In a similar fashion, for this model scenario we have shown the 1D posterior distributions of some key parameters in Fig. \ref{fig:1D-ModelII} (only for the datasets P15 and P15+GW) and in Fig. \ref{fig:1D-ModelII-A} aiming to exhibit the constraining  power of GW. 
From the analyses, we find that the simulated GWSS data are able to provide stringent constraints on the cosmological parameters, specifically, we find GWSS data very powerful to constrain some of the key cosmological parameters of the scenarios, such as the Hubble constant, $H_0$, density parameter for the matter sector, $\Omega_{m0}$, and the dark energy equation of state, $w$. However, the parameter $c_{vis}^2$ remains degenerate with every parameter of the model and this does not alter even after the inclusion of the GWSS data to the standard cosmological probes. This is pretty clear from all the plots (see for instance, Figs. \ref{fig:1D-ModelI}, \ref{fig:1D-ModelI-A}, \ref{fig:1D-ModelII}, \ref{fig:1D-ModelII-A}) shown in this article.  
This is a striking and surprising fact of this work 
where in one hand we see that GW data are extremely effective to reduce the parameter space by reducing their error bars, while on the other hand, this is not true for the viscous sound speed. We then employed the latest and final CMB data from Planck (labeled as P18) to analyse the models in order to see whether the new CMB data, i.e. P18 and also its combination with GW data, namely, P18+GW could effectively constrain the parameter $c_{vis}^2$ which has been unconstrained for the previous datasets. We found that neither the final P18 data nor the combined data P18+GW are able to constrain this parameter (see Fig. \ref{fig:1D-ModelI-Planck2018} for Model I and Fig. \ref{fig:1D-ModelII-Planck2018} for Model II). We already found that when P15 and P15+GW datasets are unable to constrain $c_{vis}^2$, the inclusion of external datasets, such as, BAO, Pantheon etc do not add any extra constraining power beyond P15 and P15+GW so that this parameter is constrained. Thus, it is clear that the inclusion of those external datasets to P18 will not be helpful in this case as well. Hence, novel and complementary astrophysical probes need to be found to probe the value of $c_{vis}^2$.

Interestingly, some complimentary probes have shown considerable progress in this direction. It Refs. \cite{Sapone:2013wda, Majerotto:2015bra} the authors took an attempt to forecast the constraints on the viscous sound speed, $c_{vis}^2$, using the Euclid and Planck surveys and found that the parameter $c_{vis}^2$ can be well constrained using those future surveys. This is certainly a fascinating report since the usual cosmological probes are unable to constrain this viscous sound speed. On the other hand, recently using the model independent approaches, the anistropic stress parameter has been confronted in presence of the  observational data \cite{Pinho:2018unz,Arjona:2020kco}. The results in  \cite{Pinho:2018unz,Arjona:2020kco} strongly suggest that a non-zero value of the anisotropic stress parameter is supported by recent observations and this consequently implies either the modification of underlying gravitational theory  or the existence of an imperfect dark energy fluid clustering at sub-horizon scales. Additionally, one can also model the anisotripc stress linked either with the DE density or DM density as in \cite{Cardona:2014iba} and within such scenarios, the underlying parameters quantifying the anisotropic stress in dark energy can be well constrained as least using the CMB data Planck as well as other cosmological probes, see \cite{Cardona:2014iba} for details.

However, the investigations are not over and the cosmic picture is still not perfectly clear. Although the future cosmological survey EUCLID \cite{Scaramella:2015rra,Laureijs:2011gra}) has played a crucial role \cite{Sapone:2013wda, Majerotto:2015bra}, however, other upcoming surveys are equally important for a better picture on the dark energy anisotropic stress. Thus, we believe that CMB Stage-4 \cite{Abitbol:2017nao}, Dark Energy Spectroscopic Instrument (DESI) \cite{Aghamousa:2016zmz}, Large Synoptic Survey Telescope (LSST) \cite{Newman:2019doi,Hlozek:2019vjs,Mandelbaum:2019zej}, Simons Observatory \cite{Ade:2018sbj}, should be employed in this framework for a clear visualization on this scenario. Such an analysis is left for a future work.

\bigskip 
\bigskip 
  
\section*{ACKNOWLEDGMENTS} 
The authors thank the referee for useful comments that helped us to improve the quality of discussion of the article. 
WY was supported by the National Natural Science Foundation of China under Grants No. 11705079 and No.  11647153. SP has been supported by the Mathematical Research Impact-Centric Support Scheme (MATRICS), File No. MTR/2018/000940, given by the Science and Engineering Research Board (SERB), Govt. of India.  DFM thanks the Research Council of Norway and the NOTUR computing facilities. MD  acknowledges the support from the National Natural Science Foundation of China under Grant No.11675032.

%-----------------------------------------------------------------

\end{document}